\documentclass[a4paper,12pt]{article}
\pdfoutput=1
\usepackage[utf8]{inputenc}
\usepackage[pdftex]{graphicx}  
\usepackage{amsfonts,amsmath,mathabx,amssymb,mathabx}
\usepackage{setspace}
 
\usepackage[top=2.7cm,left=2.7cm,right=2.7cm,bottom=2.7cm]{geometry}
\usepackage{jheppub}
\usepackage{color}
\usepackage[usenames,dvipsnames]{xcolor}
\usepackage{subfig}
\usepackage{notoccite}

\usepackage{hyperref}
\hypersetup{
    colorlinks,
    citecolor=Black,
    filecolor=Black,
    linkcolor=Black,
    urlcolor=Black
}

\usepackage{float}
\usepackage{bm} 

\numberwithin{equation}{section}
\newcommand{\be}{\begin{equation}}
\newcommand{\ee}{\end{equation}}
\newcommand{\bea}{\begin{eqnarray}}
\newcommand{\eea}{\end{eqnarray}}
\newcommand{\Tr}{{\rm Tr\,}}

\newcommand{\x}{{\rm x}}

\newcommand{\vev}[1]{{\left< {#1} \right>}}

\title{Rotating traversable wormholes in AdS}
\subheader{UTTG-11-18}
\author[a]{Elena Caceres,}
\author[a]{Anderson Seigo Misobuchi,}
\author[b]{Ming-Lei Xiao}
\affiliation[a]{Theory Group, Department of Physics, University of Texas, Austin, TX 78712, USA}
\affiliation[b]{Institute of Theoretical Physics, Chinese Academy of Science,  Beijing 100190, P. R. China}
\emailAdd{elenac@utexas.edu}
\emailAdd{anderson.misobuchi@utexas.edu}
\emailAdd{mingleix@itp.ac.cn}
\abstract{In this work we explore the effect of rotation in the size of a traversable wormhole obtained via a double trace boundary deformation. We find that  at fixed temperature {\em the size of the wormhole  increases with the angular momentum $J/M\ell$}. The amount of information that can be sent through the wormhole  increases as well. However, for the type of interaction considered, the wormhole closes as the temperature approaches the extremal limit.   We also briefly consider the scenario where the boundary coupling is not spatially homogeneous and show how this is  reflected in the wormhole opening.}

\date{\today}

\begin{document}

\maketitle

\newpage
\section{Introduction}\label{sec:intro}

Wormhole  solutions to Einstein's equations connect two asymptotically different regions of spacetime. However, they cannot be used to travel form one of the regions to the other since traversable wormholes are forbidden in classical general relativity. Their existence would require a violation of the Average Null Energy Condition (ANEC) usually achieved by including  exotic matter. ANEC states  that the stress energy tensor integrated along a complete null geodesic is always a positive quantity,  
\begin{equation}\label{eq:anec}
\int_\gamma T_{\mu \nu} k^\mu k^\nu  d\lambda \ge 0,
\end{equation}
where the null vector  $k^\mu$  is tangent to the geodesic $\gamma$  and  $\lambda $ is an affine parameter. ANEC plays a crucial  role in  singularity theorems and it has been proven to hold  \cite{Graham:2007va, Kelly:2014mra, Wall:2009wi}  along  achronal \footnote{Recall that achronal geodesics are those that do not contain any points that can be connected by a timelike curve.} null geodesics.  Null geodesics in Minkowski and in Anti-de Sitter are achronal therefore no violation of ANEC is possible and there are no traversable wormholes in neither of these spacetimes. 

In \cite{Gao:2016bin} the authors considered the scenario of an eternal black hole with two asymptotically AdS boundaries and proposed a mechanism that evades the 
assumptions of the theorems forbidding ANEC violation. They showed that, semi-classically,  including a  deformation that couples both boundaries modifies the causal structure rendering the geodesics chronal  and  ANEC can be  violated without contradicting any known theorem. Choosing an appropriate sign for the coupling we see that this is indeed the case; ANEC is violated and the wormhole becomes traversable. One can check this explicitly by following a ray coming from past infinity and traveling along the horizon. After turning on the boundary coupling this ray does not end up in the singularity but makes it  to the other boundary signaling that the wormhole has become traversable. Usually, traversable wormholes imply causal inconsistencies because one can obtain  closed time-like curves by boosting one end of the wormhole  \cite{Morris:1988tu}. The scenario of \cite{Gao:2016bin} avoids this problem since coupling both boundaries means that no such  boosts are allowed.

The scenario  presented in \cite{Gao:2016bin}   has a holographic interpretation in terms of interactions between two  CFTs and yields the first traversable wormhole solution that can be embedded  in a quantum theory of gravity. In \cite{Maldacena:2017axo} the authors elaborate on the quantum information  implications of  \cite{Gao:2016bin} and study in detail the traversable wormhole in $AdS_2$ gravity, the conjectured holographic dual of  Sachdev-Ye-Kitaev (SYK) model \cite{Sachdev:1992fk,Kitaev:2015}. They emphasize that the scenario proposed in \cite{Gao:2016bin} can be viewed as a quantum teleportation protocol. They also estimate the amount of information that can be sent through the wormhole before the wormhole closes. Other aspects of the quantum information implications of  \cite{Gao:2016bin} can be found in 
\cite{Czech:2018kvg, deBoer:2018ibj, Yoshida:2018vly, Yoshida:2017non, Almheiri:2018ijj, Susskind:2017nto, vanBreukelen:2017dul, Bak:2018txn, Miyaji:2018atq}. 

The mechanism proposed in \cite{Gao:2016bin} is beautifully simple and the result is fascinating. It could  provide an explanation of  how information can escape from a black hole. However, currently, it does not address the full black hole information paradox since for this mechanism  to work the  information should be in a very special state, the thermofield double state. There are many issues to be understood before we can  apply the traversable wormhole protocol to the black hole  information problem. Understanding the details of the traversable wormhole protocol and its quantum information implications in more general scenarios is a first step in that direction. 

In this context,  a natural question to ask is {\em how  the size of the wormhole and the amount of information transferred change  in more general gravity backgrounds}. In this work we  consider a rotating eternal black hole in three dimensions (rotating BTZ) and study the traversable wormhole produced by including a double trace deformation at the boundary. For simplicity we first consider a constant boundary coupling. We show that, at fixed temperature,  the size of the wormhole opening {\em increases} with the angular momentum.   We establish a  bound on information that can be transferred and show that in a rotating background more information can be sent through the wormhole as compared with the non-rotating scenario of \cite{Gao:2016bin} and \cite{Maldacena:2017axo}. We show that our results are valid for relevant operators of any conformal dimension $(0<\Delta< 1)$. The increase of the bound  is particularly noticeable at higher angular momentum $J$.  We also analyze the extremal limit and find that the wormhole closes as we approach  $J=M$.
The couplings considered in \cite{Gao:2016bin} and \cite{Maldacena:2017axo} were homogeneous in the  boundary spatial directions. We briefly consider the effect of taking a coupling  with dependence on the $x$ boundary directions  and investigate how this affects the wormhole opening. 

This paper is organized as follows: in section \ref{sec:review} we review the rotating BTZ geometry and establish our notation. In sections \ref{sec:opening} and \ref{sec:bound} we explore ANEC violation, the size of the traversable wormhole and the bound on information that can be transferred through it; these sections contain our main results. In section \ref{sec:transverse} we briefly consider the case of a non-homogeneous coupling. Section \ref{sec:conclusions} contains a summary of our results and interesting future directions.

\section{Review of the rotating BTZ geometry}\label{sec:review}
The rotating BTZ black hole \cite{Banados:1992gq,Banados:1992wn} is a solution of Einstein gravity in $2+1$ dimensions with negative cosmological constant $\Lambda=-1/\ell^2$ described by the action
\begin{equation}
I=\frac{1}{16\pi G_N}\int d^3x\sqrt{-g}(R+2\ell^{-2}).
\end{equation}
The solution can be constructed from a quotient of global $AdS_3$. The metric in $(t,\ r,\ \tilde{x})$ coordinates is,
\begin{align} \label{metric}
 & ds^2=-\frac{(r^2-r_+^2)(r^2-r_-^2)}{\ell^2 r^2}dt^2+\frac{\ell^2 r^2}{(r^2-r_+^2)(r^2-r_-^2)}dr^2+r^2\left(d\tilde{x}-\frac{r_+r_-}{\ell r^2}dt\right)^2, \nonumber\\
 & \tilde{x}\sim\tilde{x}+2\pi. 
\end{align}

The inner (Cauchy) horizon is $r_-$ and the outer horizon is $r_+$. The identification in the angular coordinate $\tilde{x}$ breaks the global $SL(2,\mathbb{R})\times SL(2,\mathbb{R})$ isometry of $AdS_3$ down to a $\mathbb{R}\times SO(2)$ subgroup. Without the identification, the solution is simply a portion of $AdS_3$ with a Rindler horizon for an accelerated observer. Since our interest is the region near the outer horizon, it is convenient to work in the co-rotating frame with the shifted angular coordinate
\begin{equation}
 x\equiv\tilde{x}-\Omega_H t,
\end{equation}
where $\Omega_H=\frac{r_-}{\ell r_+}$ is the angular velocity of the outer horizon. With this choice the metric becomes
\begin{align}
 & ds^2=-\frac{(r^2-r_+^2)(r^2-r_-^2)}{\ell^2 r^2}dt^2+\frac{\ell^2 r^2}{(r^2-r_+^2)(r^2-r_-^2)}dr^2+r^2\left(\mathcal{N}(r)dt+dx\right)^2,\nonumber \\
 & \mathcal{N}(r)=\frac{r_-}{2r_+}\frac{r^2-r_+^2}{\ell r^2},\qquad x\sim x+2\pi.
\end{align}
This choice is useful because the off-diagonal terms in the metric vanish at the outer horizon. 

The thermodynamic variables are,
\begin{equation} \label{eq:thermo}
 M=\frac{r_+^2+r_-^2}{8 G_N\ell^2}, \quad J=\frac{r_+r_-}{4G_N \ell},\quad S=\frac{\pi r_+}{2 G_N}, \quad \kappa=\frac{r_+^2-r_-^2}{\ell^2 r_+}, \quad\beta=\frac{2\pi}{\kappa},
\end{equation}
where $M$ is the mass, $J$ is the angular momentum, $S$ is the entropy, and $\kappa$ is the surface gravity of the outer horizon, related to the temperature of the black hole by $\kappa=2\pi/\beta=2\pi T$. The horizon exists provided $M>0$ and $|J|\leq M\ell$. Without loss of generality we can assume that $J$ is positive. The extremal limit, in which the two horizons coincide, corresponds to $|J|=M\ell$. We can also rewrite the inner and outer horizons in terms of $M$ and $J$
\begin{equation} \label{rprm}
 r_\pm^2=\frac{1}{16 G_N}\left(M\ell^2\pm\sqrt{(M\ell^2)^2-J^2\ell^2}\right).
\end{equation}

\begin{figure}[h!]
\center
\subfloat[]{\includegraphics[height=0.4\textheight]{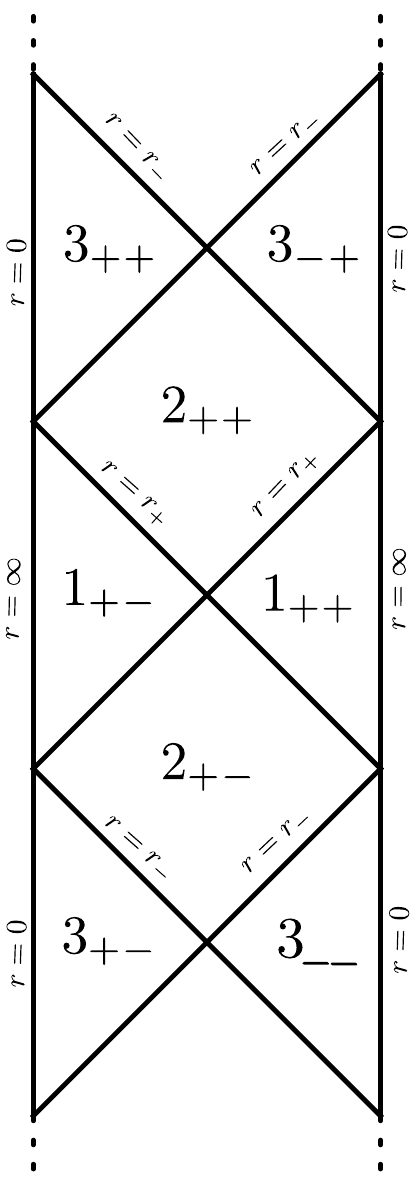}}
\qquad\qquad
\subfloat[]{\includegraphics[height=0.4\textheight]{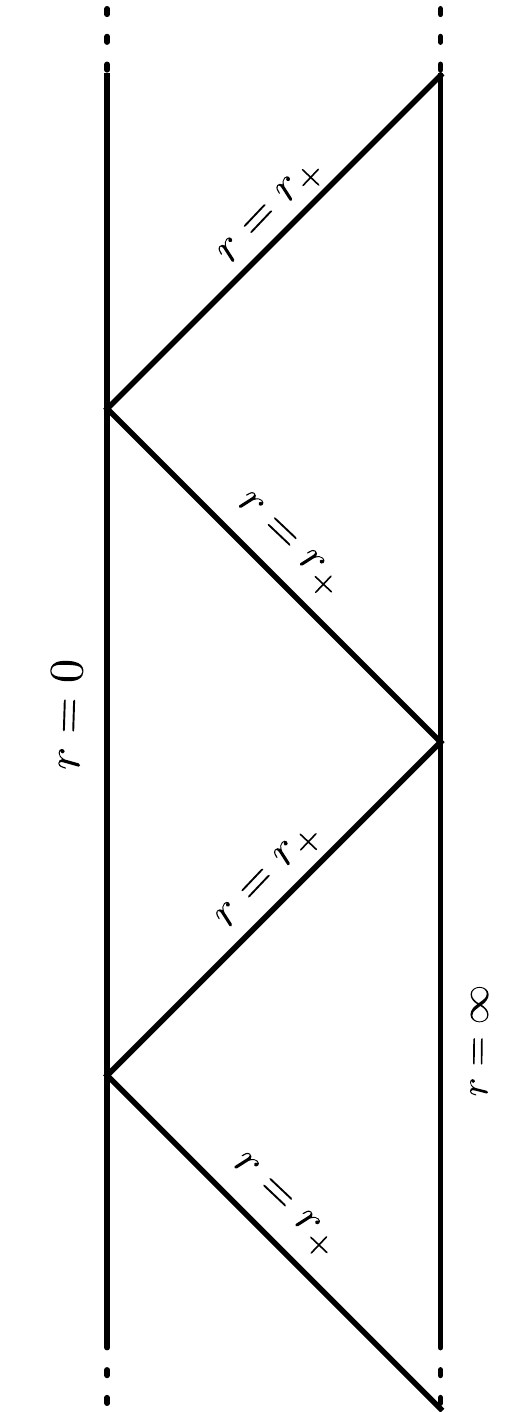}}
\caption{(a) Penrose diagram for maximally extended rotating BTZ (non-extremal). Regions $1_{+\pm}$: $r>r_+$. Regions $2_{+\pm}$: $r_-<r<r_+$. Regions $3_{\pm\pm}$: $r<r_-$. (b) Penrose diagram for the extremal limit $r_+=r_-$.}
\label{penrose}
\end{figure}

The Penrose diagram for the maximal extension of the rotating BTZ black hole is depicted in Fig.\,\ref{penrose}. In principle, the maximal extension allows to continue to negative values of $r^2$, but since this introduces closed timelike curves the diagram is truncated at the surface $r=0$, which is treated as a singularity (see \cite{Hemming:2002kd} for a more detailed discussion). Note that we will not concern ourselves with the regions behind the $r_-$ horizon ($3_{++}, 3_{-+}, 3_{+-}, 3_{--})$ since the $r_-$  horizon is unstable \cite{Balasubramanian:2004zu}. Regions $1_{++}$ and $1_{+-}$ correspond to the region outside the horizon ($r_+<r<\infty$). Embedding coordinates appropriate for region $1_{++}$ are given by \cite{Balasubramanian:2004zu}
\begin{equation}
1_{++}:
 \begin{cases}
 X_+=\left[\left(\frac{r-r_+}{r+r_+}\right)\left(\frac{r+r_-}{r-r_-}\right)^{r_-/r_+}\right]^\frac{1}{2} \cosh(\kappa \,t)\\
 T_+=\,\,\left[\left(\frac{r-r_+}{r+r_+}\right)\left(\frac{r+r_-}{r-r_-}\right)^{r_-/r_+}\right]^\frac{1}{2} \sinh(\kappa \,t)\\
 \end{cases},\quad r_+<r<\infty.
\end{equation}
We can reach the region $1_{+-}$ from $1_{++}$ by performing a imaginary shift in time $t\to t-i\beta/2$. Similarly, regions $2_{++}$ and $2_{+-}$ can be reached from $1_{++}$ shifting time by  $-i\beta/4$ and $-3i\beta/4$, respectively\footnote{In the co-rotating frame there is no change in $x$ when we move to other region, but in the original coordinate system $\tilde{x}$ is shifted by $\tilde{x}\to \tilde{x}-i\beta\Omega_H/2$ when we move a point from $1_{++}$ to $1_{+-}$, for example.}. Kruskal coordinates are defined as
\begin{equation} \label{kruskal}
 U=X_++T_+,\qquad V=X_+-T_+.
\end{equation}
This choice of Kruskal coordinates is valid only for $r>r_-$. The metric \eqref{metric} becomes
\begin{align} \label{eq:metric_kruskal}
 & ds^2=-\Omega^2(r)dUdV+\left(\frac{\mathcal{N}(r)r}{2U\kappa}dU-\frac{\mathcal{N}(r)r}{2V\kappa}dV+r dx\right)^2,\nonumber \\ 
 &\Omega^2(r)\equiv\frac{(r^2-r_-^2)(r+r_+)^2}{\kappa^2 r^2\ell^2}\left(\frac{r-r_-}{r+r_-}\right)^{r_-/r_+},
\end{align}
where $r=r(U V)$ is understood as an implicit function. Note that the coordinate transformation becomes singular when $r_+=r_-$, so taking the strict limit $r_+=r_-$ is not possible with this choice of coordinates. For this reason, the extremal limit in our analysis will correspond to approach $r_+\to r_-$, but still keeping $r_+\neq r_-$.

\section{Opening the wormhole with a double trace deformation}\label{sec:opening}

In AdS/CFT the eternal AdS black hole geometry is understood to be dual to two copies of a CFT living in the two asymptotic boundaries (corresponding to regions $1_{++}$ and $1_{+-}$ in Fig.\,\ref{penrose}) in the thermofield double state (TFD) \cite{Maldacena:2001kr},
\begin{equation}
 |\Psi\rangle=\frac{1}{\sqrt{Z(\beta,\Omega_H)}}\sum_{n} e^{-\beta(E_n-\Omega_H J_n)/2}|E_n,J_n\rangle_L|E_n,J_n\rangle_R,
\end{equation}
where $Z=\Tr e^{-\beta(H-\Omega_H J)}$ is the partition function. The sum runs over a complete basis of eigenstates with energy $E_n$ and angular momentum $J_n$.  We can view the combination $\tilde{H}_{L/R}=H_{L/R}-\Omega_H J_{L/R}$  as the effective Hamiltonian that evolves the Left/Right system. When we consider the evolution for the full system defined in the Hilbert space $\mathcal{H}=\mathcal{H}_L\otimes\mathcal{H}_R$ there are two possibilities for the choice of sign for the Hamiltonian. Here we choose the combination $\tilde{H}_R-\tilde{H}_L$ which leaves the TFD invariant.

In the identification of the eternal rotating BTZ with the TFD state, it is implicitly  assumed that the dual theory is  defined in the two boundaries in regions $1_{++}$ and $1_{+-}$ instead of the infinite disconnected boundaries present in the maximal extension \cite{Krishnan:2009kj,Levi:2003cx}. Moreover, as previously mentioned,  the region behind the $r_-$ horizon  is  excluded since it is known that scalar perturbations lead to instabilities of the Cauchy horizon. In our work the focus is in the region close to  the outer horizon so this subtlety plays no role in our analysis.

Gao, Jafferis and Wall (GJW) \cite{Gao:2016bin} showed that a wormhole in the eternal $AdS$ black hole scenario can be made traversable by turning on a coupling between the left and right boundaries of form
\begin{equation}\label{interaction}
\delta H(t_1)=-\int\,d^{d-1}x_1\,h(t_1,x_1)\mathcal{O}_R(t_1,x_1)\mathcal{O}_L(-t_1,x_1), 
\end{equation}
where for simplicity we choose,
\begin{equation}
h(t_1,x_1)=
\begin{cases} h\kappa^{2-2\Delta}, \quad t_0\leq t_1\leq t_f \\0\qquad \quad, \quad \text{otherwise.} \end{cases}
\end{equation}
$\mathcal{O}_{L/R}$ is a scalar operator of dimension\footnote{In principle there are two choices of sign $\Delta_\pm=\frac{d}{2}\pm\sqrt{\left(\frac{d}{2}\right)^2+m^2}$, but we pick the minus sign to have a relevant deformation, which constrains $0<\Delta<1$ for $d=2$.} $\Delta=\frac{d}{2}-\sqrt{\left(\frac{d}{2}\right)^2+m^2}$ living in the Left/Right CFT, dual to a bulk scalar field $\Phi$ with mass $m$. For a suitable choice of the sign of the coupling this interaction produces negative null energy in the bulk violating the Averaged Null Energy Condition (ANEC). Using the linearized Einstein equations we will show in Sec.\,\ref{sec:wormhole_size} that the effect of the interaction is to produce a negative shift $\Delta V$, so that a test particle sent from one boundary traveling near the horizon can reach the other side.

\subsection{Bulk-to-boundary propagator in a rotating background }

One key ingredient to study traversability is the knowledge of the bulk-to-boundary propagator. Since the rotating BTZ solution is locally $AdS_3$, the bulk-to-boundary propagator can be obtained from the propagator for $AdS_3$ via a coordinate transformation. The result when both bulk and boundary points are contained in region $1_{++}$ is \cite{Balasubramanian:2004zu},
\begin{align} \label{prop}
 & \mathcal{K}(z,t,x;t_1,x_1)=\nonumber\\
 & \quad \frac{(r_+^2-r_-^2)^\frac{\Delta}{2}}{2^{\Delta+1}\pi \ell}\sum_{n={-\infty}}^{\infty}\left[-\sqrt{z-1}\cosh\left(\kappa\delta t-\frac{r_-}{\ell}\delta x_n\right)+\sqrt{z}\cosh\left(\frac{r_+}{\ell}\delta x_n\right)\right]^{-\Delta},
\end{align}
where
\begin{equation}
 z=\frac{r^2-r_-^2}{r_+^2-r_-^2},\quad \delta t=t-t_1, \quad \text{and} \quad\delta x_n=x-x_1+2\pi n.
\end{equation}

Note that the overall normalization of the propagator is obtained by taking the limit to the boundary from the bulk-to-bulk propagator \cite{Ichinose:1994rg,Azeyanagi:2007bj}, such that it agrees with the propagator considered by GJW in the limit $r_-\to0$.
Near the event horizon in the region $1_{++}$, it is possible to explicitly invert the Kruskal coordinates \eqref{kruskal} to obtain
\begin{equation}
 t=\frac{1}{2\kappa}\log\left(-\frac{U}{V}\right), \qquad z= 1-\gamma^2 \,UV +O(U^2V^2),
\end{equation}
where we have defined 
\begin{equation} \label{gamma}
 \gamma^2\equiv\Omega^2(r=r_+)=\frac{4r_+^2\ell^2}{r_+^2-r_-^2}\left(\frac{r_+-r_-}{r_++r_-}\right)^{r_-/r_+}.
\end{equation}
For example, the bulk-to-boundary propagator along $V=0$ for both points contained in region $1_{++}$ becomes (omiting sum over images)
\begin{equation}
 \mathcal{K}(U,0,x;\,U_1,x_1)=\frac{\left(r_+^2-r_-^2\right)^{\frac{\Delta}{2}}}{2^{\Delta+1}\pi\ell}\left(\frac{1}{-\frac{\gamma}{2} U/U_1 e^{-r_-(x-x_1)}+\cosh[r_+(x-x_1)]}\right)^\Delta,
\end{equation}
and propagators for points in different regions can be obtained using the imaginary shifts in time that we have described in Sec.\,\ref{sec:review}. We will also need the retarded bulk-to-boundary propagator, which can be expressed as
\begin{align}\label{retarded}
 &\mathcal{K}_\text{ret}(z,t,x;t_1,x_1)=\\
 &= |\mathcal{K}(z,t,x;t_1,x_1)|\,\theta(\delta t)\,\theta\left(\sqrt{z-1}\cosh\left(\kappa\,\delta t-\frac{r_-}{\ell}\delta x\right)-\sqrt{z}\cosh\left(\frac{r_+}{\ell}\delta x\right) \right).\nonumber
\end{align}
Throughout the rest of the paper we will set $\ell=1$ for simplicity.

\subsection{ANEC violation and wormhole size}\label{sec:wormhole_size}

We now evaluate the modified stress tensor in the rotating BTZ background when we turn on the interaction \eqref{interaction}. The analytic continuation of the bulk-to-boundary propagator when the points are time-like separated from the boundary was studied in  \cite{Levi:2003cx}. This continuation works very much like in the non-rotating case. Therefore, the steps of the calculation follow closely  \cite{Gao:2016bin}.

The starting point is to evaluate the bulk two-point function 
\begin{equation}
 G(U,U')\equiv \vev{\Phi_R(U,x)\Phi_R(U',x)}.
\end{equation}
In the perturbative expansion in the coupling $h$, the one-loop contribution to the two-point function is
\begin{equation}
 G_h= \frac{2 h \sin(\pi\Delta)}{\kappa^{2\Delta-2}}\int_{t_0}^{t_f} dt_1 dx_1\mathcal{K}(r',t',x';-t_1+i\tfrac{\beta}{2},x_1)\mathcal{K}_\text{ret}(r,t,x;t_1,x_1)+(t\leftrightarrow t'),
\end{equation}
where we used the imaginary shift $t\to t-i\beta/2$ to move all the points to the right region $1_{++}$. Using the propagators \eqref{prop} and \eqref{retarded}, and evaluating at $V=0$ in Kruskal coordinates gives
\begin{align}
 & G_h(U,U')=\nonumber\\
 &C_0\int_{U_0}^U \frac{dU_1}{U_1}\left(\frac{1}{\frac{\gamma}{2}e^{-r_-\delta x}U_1 U'+\cosh(r_+\delta x)}\right)^\Delta \left(\frac{U_1}{\frac{\gamma}{2}e^{-r_-\delta x}U-U_1\cosh(r_+\delta x)}\right)^\Delta dx_1\nonumber \\
 &\hspace{5in}+(U\leftrightarrow U') \label{eq:pert_prop}
\end{align}
where $U_0=e^{\kappa \,t_0}$. After we shut down the interaction ($U>U_f=e^{\kappa\, t_f}$) the upper limit in the $U_1$ integral should be replaced by $U_f$. The overall constant is
\begin{equation}
 C_0=\frac{h\kappa^{1-\Delta}r_+^\Delta \sin(\Delta\pi)}{2(2^\Delta\pi)^2}.
\end{equation}
The sum over the images extends the domain of the $x_1$ integral to the entire real axis, but it ends being constrained by the $\theta-$function in the retarded propagator \eqref{retarded}, which requires that
\begin{equation}
 \frac{\gamma}{2}\,e^{-r_-\delta x}U-U_1\cosh(r_+\delta x) \geq 0.
\end{equation}
The bulk stress tensor associated to the scalar field $\Phi$ is
\begin{equation}
 T_{\mu\nu}=\partial_\mu\Phi\partial_\nu\Phi-\frac{1}{2}g_{\mu\nu}g^{\rho\sigma}\partial_\rho\Phi\partial_\sigma\Phi-\frac{1}{2}g_{\mu\nu}m^2\Phi^2.
\end{equation}
At one-loop, the expectation value of the stress tensor for the perturbed Hamiltonian can be evaluated via point splitting
\begin{equation}
 \vev{T_{\mu\nu}}=\lim_{\x\to \x'}\left(\partial_\mu\partial_\nu G(\x,\x')-\frac{1}{2}g_{\mu\nu}g^{\rho\sigma}\partial_\rho\partial_\sigma G(\x,\x')-\frac{1}{2}g_{\mu\nu}m^2 G(\x,\x')\right).
\end{equation}
When evaluated along the horizon at $V=0$, the $g_{UU}$ component of the unperturbed metric vanishes, so the leading contribution to the null component of the stress tensor is,
\begin{equation}\label{eq:TUU_pert}
 T_{UU}=\lim_{U'\to U}\partial_U\partial_{U'}G_h (U,U').
\end{equation}
where the perturbed propagator $G_h(U,U')$, obtained  in \eqref{eq:pert_prop}, can be evaluated numerically.  Evaluating \eqref{eq:TUU_pert} we can   find  the effect of the double trace deformation on the average null energy.  As we see in Fig.\,\ref{fig:ANEC} our results show that the ANEC is violated. 

As mentioned in Sec.\,\ref{sec:intro}, ANEC violation is a necessary condition for a wormhole becoming traversable. It is illuminating to explicitly follow a null ray in the perturbed metric  going  from past infinity ($U\rightarrow -\infty$) to future infinity ($U\rightarrow \infty$)  along the $V=0$ horizon.
The linearized Einstein equation for the $UU$ component for the fluctuations evaluated at $V=0$ gives 
\begin{equation}
\int dU\left(\frac{\kappa}{2r_+}h_{UU}-\frac{r_- \partial_x h_{UU}}{r_+^2}-\frac{\partial_x^2 h_{UU}}{2r_+^2}\right)=8\pi G_N \int dU T_{UU},
\end{equation}
where we assume that the fluctuations vanish at infinity. Since the interaction \eqref{interaction} is being integrated over the transverse space, the modified stress tensor should be independent of the transverse coordinate $x$. This reduces the equation to
\begin{equation} \label{linearized}
 8\pi G_N\int dUT_{UU}=\frac{\kappa}{2 r_+}\int dU h_{UU}=\frac{r_+^2-r_-^2}{2r_+^2}\int dU h_{UU}.
\end{equation}
 We can now relate the average null energy to the shift $\Delta V(U)$ in the null geodesics at the horizon caused by the interaction. After including the perturbation, we can see from the metric \eqref{eq:metric_kruskal} that the null ray originating in the past is given by, 
\begin{align}\label{eq:null_ray}
 \Delta V(U)& =-\frac{1}{2g_{UV}(0)}\int_{-\infty}^{U}dU h_{UU}
 \end{align}
where $g_{UV}(0)$ is the $U\,V$ component of the original metric evaluated on $V=0$. Now, from \eqref{linearized} and \eqref{eq:null_ray} we get,
  \begin{equation} \label{eq:DeltaV} 
  \Delta V(U)=\frac{1}{2}\left(\frac{r_+-r_-}{r_++r_-}\right)^{-\frac{r_-}{r_+}}8\pi G_N\int_{-\infty}^{U}dUT_{UU}.
\end{equation}
We identify the ``size'' of opening of the wormhole as $\Delta V(\infty)$, which we will denote simply by $\Delta V$. 

\subsubsection{Numerical results} \label{sec:numerics}

\begin{figure}[h!]
	\center
	\includegraphics[width=0.8\linewidth]{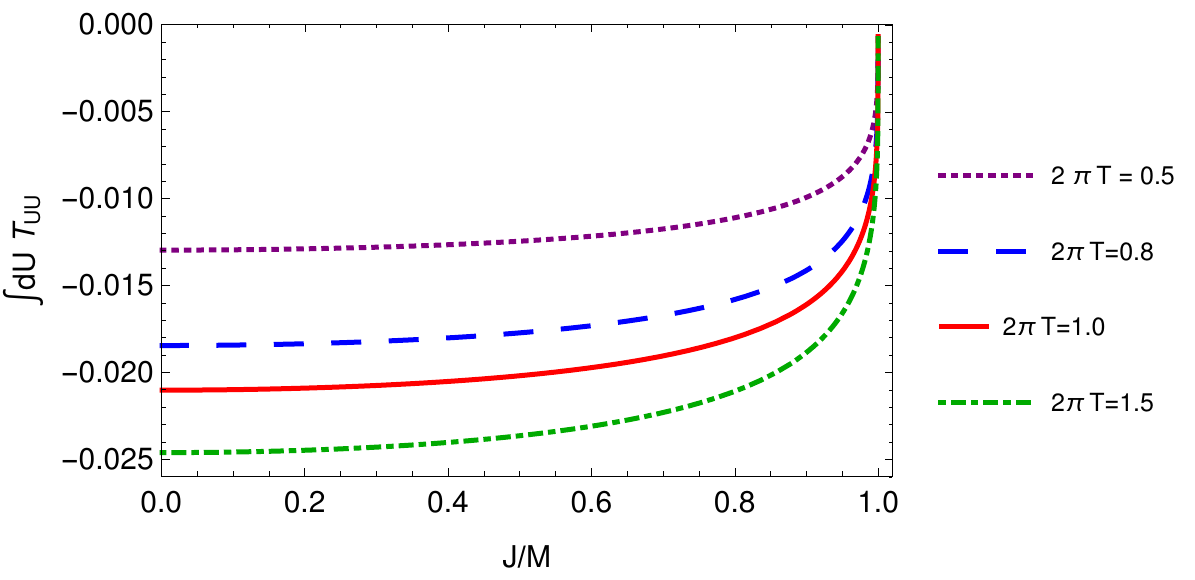}
	\caption{Average null energy at fixed temperatures as a function of angular momentum. We have set $\Delta=0.6$. }
	\label{fig:ANEC}
\end{figure}

In our numerical analysis we have set $h=1$, $G_N=1$, and the interaction was turned on between $t_0=0$ and $t_f=1$. Fig.\,\ref{fig:ANEC} shows the ANE obtained by evaluating \eqref{eq:TUU_pert} numerically at fixed temperatures as a function of the dimensionless ratio $J/M$. Note that the curves never reach the value $J/M=1$ since we are fixing a non-zero temperature (non-extremal case). We see that the ANEC is violated and this violation is more pronounced for increasing temperatures. As a check of our numerical results, note that the red (solid) curve in Fig.\,\ref{fig:ANEC}  corresponds to the same temperature chosen in \cite{Gao:2016bin} and the value for $J=0$ in our plot coincides with the numerical value they obtained for $\Delta=0.6$.

\begin{figure}[h!]
	\center
	\includegraphics[width=0.8\linewidth]{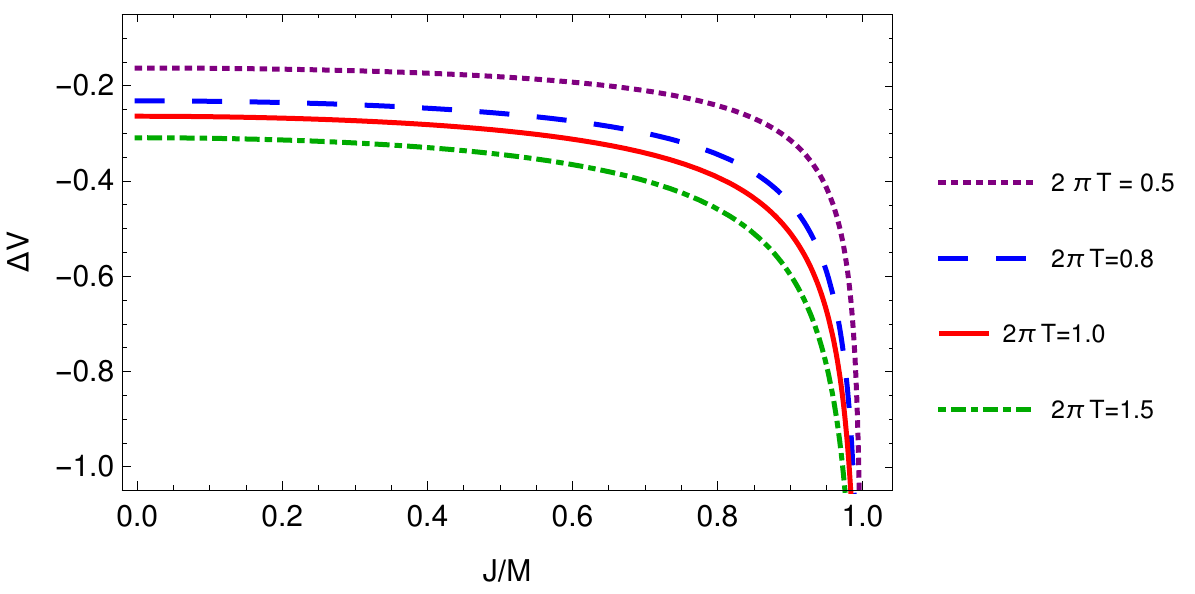}
	\caption{Size of the opening of the wormhole as a function of angular momentum. We have set $\Delta=0.6$.}
	\label{fig:sizeJM}
\end{figure}

One of the motivations of this work is to find out if the traversable wormhole is larger in a rotating background. We find that the answer is positive for $h>0$. 
{\em The wormhole opening becomes larger as we increase $J/M$}; see Fig.\,\ref{fig:sizeJM}. This increasing is mainly due to the geometrical factor in \eqref{eq:DeltaV} relating the opening of the wormhole and the ANE. In the next section we will re-derive this result using the out-of-time order correlators approach developed in \cite {Maldacena:2017axo}.

\begin{figure}[h!]
	\center
	\subfloat[Fixed $M=1$.\label{fig:opextremal_M}]{\includegraphics[width=0.45\linewidth]{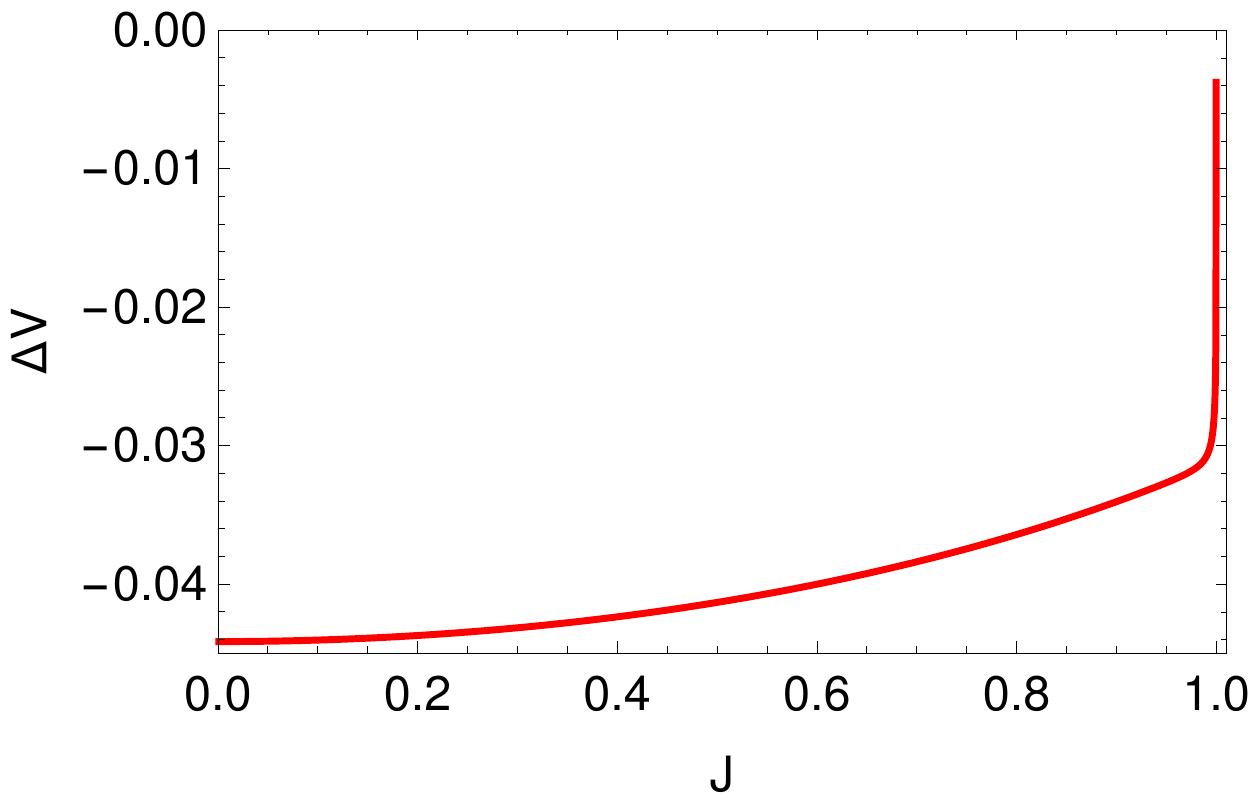}}
	\qquad
	\subfloat[Fixed $r_+=1$.\label{fig:opextremal_rp}]{\includegraphics[width=0.45\linewidth]{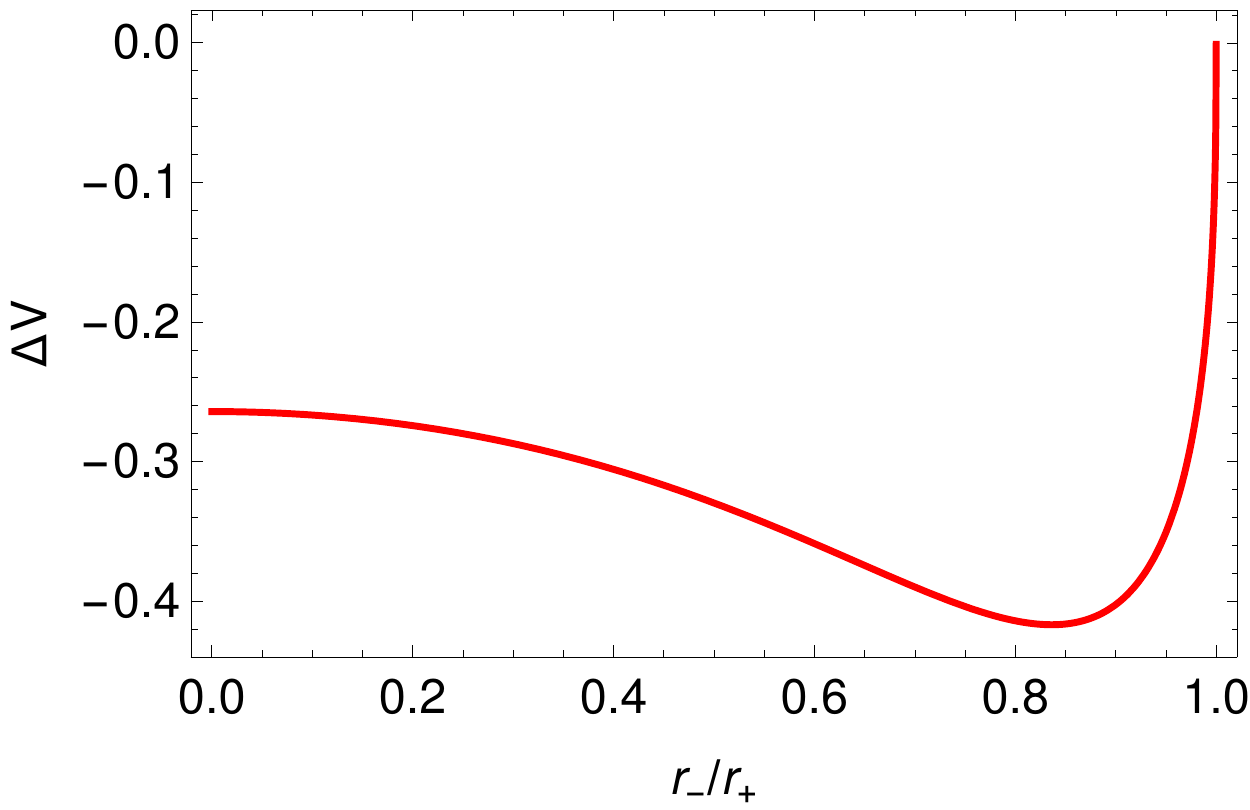}}
	\caption{Opening of the wormhole with temperature not fixed. The wormhole closes as we approach the extremal limit. We have set $\Delta=0.6$.}
	\label{fig:opextremal} 
\end{figure}

So far we have only considered the case of fixed temperatures that does not give us much information about the extremal limit, in which $T\to0$ and $J\to M$. We can investigate the extremal limit by keeping the mass fixed and varying $J$ and $T$ simultaneously. Our results show that the wormhole closes as we approach the extremal limit; see Fig.\,\ref{fig:opextremal_M}. We note that the variation in the opening near the extremal limit is very abrupt if we take the limit with fixed mass. We have also plotted the opening by fixing $r_+$, which corresponds to fix the black hole entropy, as a function of the angular velocity $\Omega_H=\frac{r_-}{r_+}$; see Fig.\,\ref{fig:opextremal_rp}.

\begin{figure}[h!]
	\center
{\includegraphics[width=0.8\linewidth]{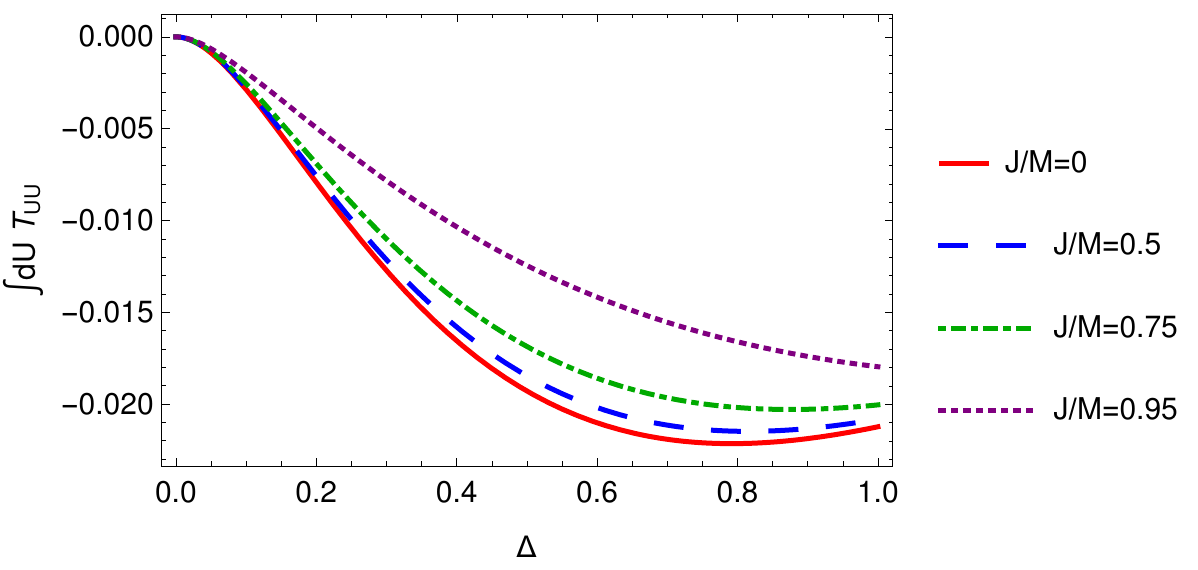}}
	\caption{Average null energy for different values of the conformal dimension $\Delta$ at fixed temperature $T=\frac{1}{2\pi}$.}
\label{fig:ANE_different_deltas} 
\end{figure}

\begin{figure}[h!]
	\center
{\includegraphics[width=0.8\linewidth]{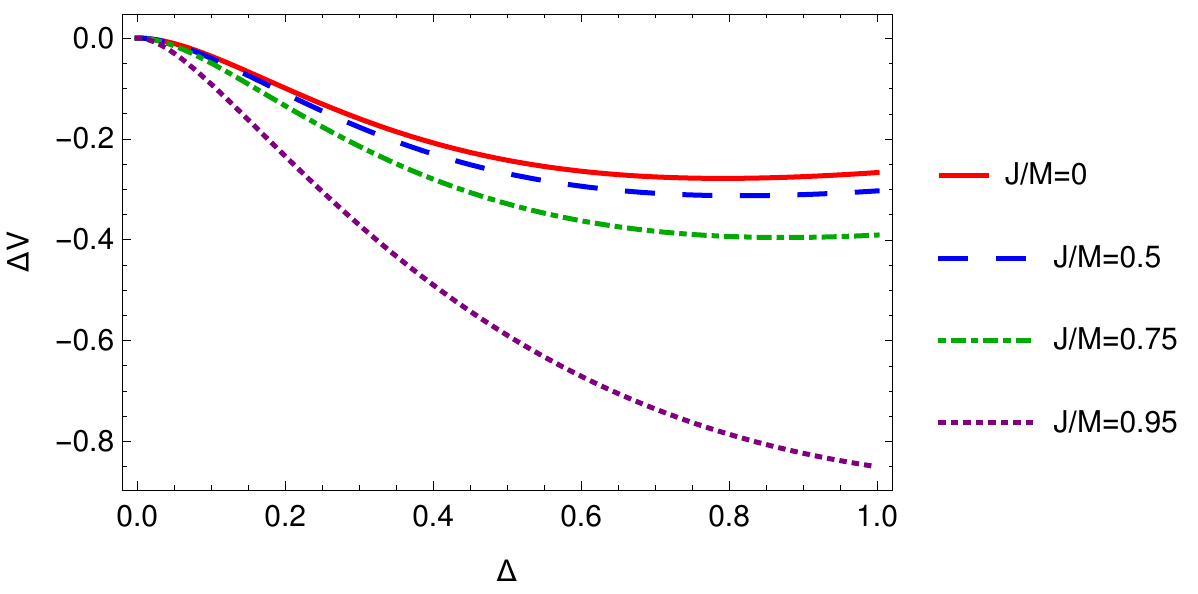}}
	\caption{Wormhole opening for different values of the conformal dimension $\Delta$ at fixed temperature $T=\frac{1}{2\pi}$.}
	\label{fig:size_different_deltas} 
\end{figure}

The numerical results presented in Fig.\,\ref{fig:ANEC}-\ref{fig:opextremal} were obtained for  boundary operators  of conformal dimension $\Delta=0.6$. We repeat the calculation for different values of $\Delta$ and we find that for any value of allowed conformal dimension ($0<\Delta<1$) there is a violation of ANEC implying a traversable wormhole; see Fig.\,\ref{fig:ANE_different_deltas}-\ref{fig:size_different_deltas}.

\section{Bound on information transfer and backreaction}\label{sec:bound}

\subsection{Diagnose of traversability from Left/Right commutator}

\begin{figure}[h!]
	\center
	\includegraphics[width=0.6\linewidth]{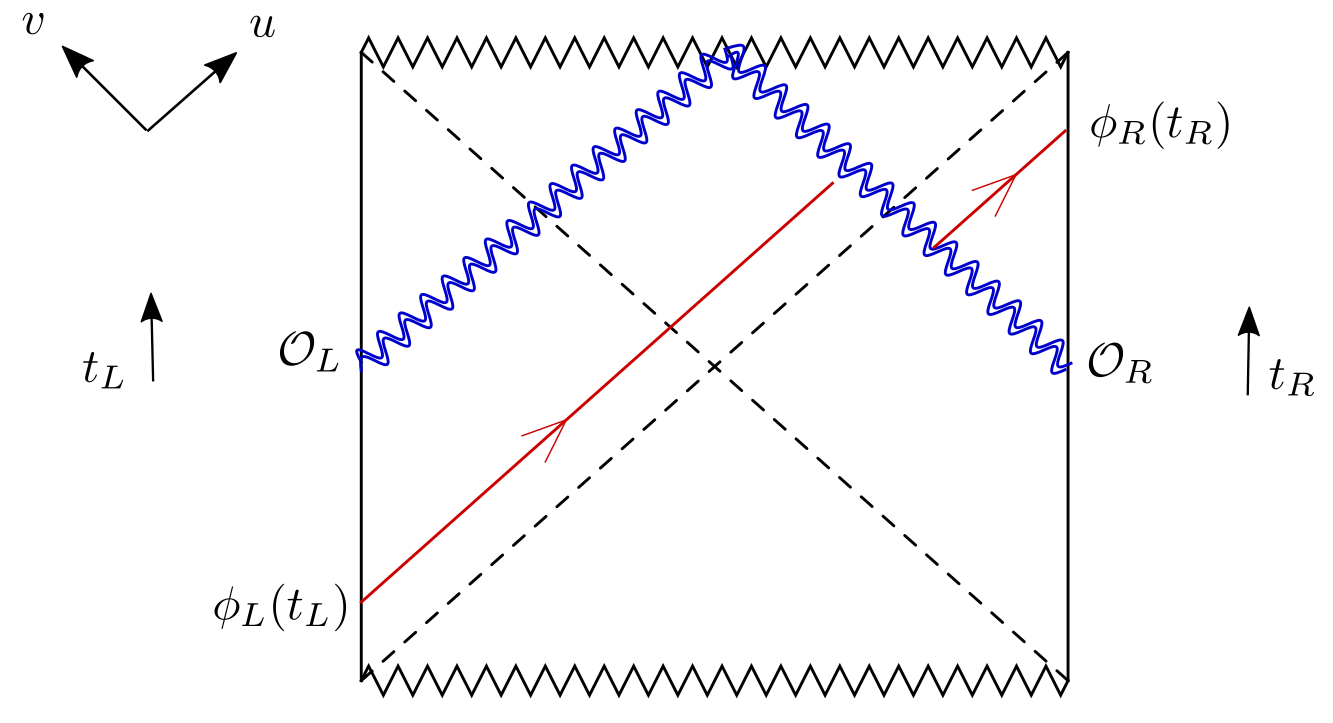}
	\caption{The setup of the traversable wormhole \cite{Maldacena:2017axo}. The double trace deformation produces negative null energy that causes a time advance in the probe $\phi$.}
	\label{fig:setup}
\end{figure}

In \cite{Maldacena:2017axo} the authors elaborated on the quantum information interpretation of the results in \cite{Gao:2016bin}. Focusing on nearly $AdS_2$ gravity \cite{Maldacena:2016upp}, which is claimed to be the dual of the SYK model, they presented a bound on the information that can be transferred through the wormhole. They also considered a double trace deformation\footnote{More precisely, they considered the slightly different interaction $\frac{g}{K}\sum_{j=1}^K\mathcal{O}^j_R(0)\mathcal{O}^j_L(0)$, where the $K$ fields are introduced to make the effect larger and the interaction is turned on only at time $t=0$.} but proposed a different way to diagnose traversability of the wormhole using techniques involving out-of-time order correlation functions that appear in the context of quantum chaos \cite{Shenker:2013pqa, Shenker:2014cwa}. In this perspective, traversability can be interpreted as the result of a high energy scattering where the signal particle $\phi$ scatters with $\mathcal{O}$ and suffers a time advance for positive $h$, so it can emerge on the other boundary (Fig.\,\ref{fig:setup}).

In this section we review the diagnose of \cite{Maldacena:2017axo} and we adapt their results to derive the opening of the wormhole for the eternal rotating BTZ black hole. We consider a generic metric in Kruskal coordinates written as
\begin{equation} \label{eq:metric_generic}
 ds^2=-a(uv)dudv+\left(b(u,v)du+c(u,v)dv+r(uv)dx\right)^2, 
\end{equation}
and let us define the following quantities evaluated at the horizon 
\begin{equation} \label{eq:a0}
\quad a_0=a(0),\quad r_0=r(0).
\end{equation}
We will assume that $b(v=0)=0$ to simplify the calculation. Note that this applies to the rotating BTZ case \eqref{eq:metric_kruskal}. In the framework of \cite{Maldacena:2017axo} the diagnose of traversability is based on the commutator
\begin{equation}
 \vev{\left[\phi_L, e^{-i\mathcal{V}}\phi_R\, e^{i\mathcal{V}}\right]}, \qquad \mathcal{V}\equiv\int_{t_0}^{t}dt_1\,\delta H(t_1),
\end{equation}
where $\phi_R=\phi_R(t_R)$, $\phi_L=\phi_L(t_L)$, and the brackets $\vev{...}$ denote the expectation value in the TFD state. The convention for time is $t_R=-t_L=t$. If the commutator is non-zero, then the signal from the left boundary successfully reached the right boundary.

Assuming the operators are Hermitian, we can focus on a simpler quantity
\begin{equation}
 C\equiv \vev{e^{-i\mathcal{V}}\phi_R\,e^{i\mathcal{V}}\phi_L},
\end{equation}
which is related to the original commutator via
\begin{equation}
 \vev{\left[\phi_L, e^{-i\mathcal{V}}\phi_R\,e^{i\mathcal{V}}\right]}=-2\,\text{Im}(C).
\end{equation}
In the small $G_N$ limit, and assuming the relative boost between $\mathcal{O}$ and $\phi$ is large, we can approximate the scattering between states created by $\phi$ and $\mathcal{O}$ by a shock wave with amplitude $S_{\text{grav}}=e^{i\tilde{\delta}}$ \cite{Verlinde:1991iu,Kabat:1992tb}. The final expression is \cite{Maldacena:2017axo} (see also \cite{Almheiri:2018ijj})
\begin{equation} \label{eqC}
 C = e^{i\vev{\mathcal{V}}}\alpha\int dp^u\,dy\, p^u\Psi_{\phi_L}^*(p^u,y)\Psi_{\phi_R}(p^u,y)\,e^{i\mathcal{D}},
\end{equation}
where $\alpha=\frac{a_0^2r_0}{4\pi}$. The exponent is defined by
\begin{equation} \label{eqD}
 \mathcal{D}\equiv\alpha \int dq^v\,dx \int dt_1\,dx_1\,q^v e^{i\tilde{\delta}}h(t_1,x_1)\Psi_{\mathcal{O}_R}^*(q^v,x)\Psi_{\mathcal{O}_L}(q^v,x),
\end{equation}
where the wavefunctions are given by the Fourier transform of the bulk-to-boundary propagators
\begin{align}
 \Psi_{\mathcal{O}_R}(q^v,x) & =\int dv e^{ia_0 q^v v/2}\vev{\Phi(u,v,x)\,\mathcal{O}_R^\dagger(t_1,x_1)}_{u=0}, \nonumber\\
 \Psi_{\mathcal{O}_L}(q^v,x) & =\int dv e^{ia_0 q^v v/2}\vev{\Phi(u,v,x)\,\mathcal{O}_L(-t_1,x_1)}_{u=0}, \nonumber\\
 \Psi_{\phi_R}(p^u,y) &= \int du e^{ia_0 p^u u/2}\vev{\Phi_\phi(u,v,y)\,\phi_R(t,y_1)}_{v=0},\nonumber \\
 \Psi_{\phi_L}(p^u,y) &=\int du e^{ia_0 p^u u/2}\vev{\Phi_\phi(u,v,y)\,\phi_L^\dagger(-t,y_1)}_{v=0},
\end{align}
and $\Phi_\phi$ is the bulk field dual to $\phi$. In the above expression, we have decomposed the single particle wavefunction of $\phi$ into states with momentum $p^u$ on the $v=0$ slice, and momentum $q^v$ on the $u=0$ slice for $\mathcal{O}$. The operator $\mathcal{O}_{L/R}$ is applied at a point $x_1$ and $\phi_{L/R}$ is applied at $y_1$ in transverse space, while $x$ and $y$ are bulk transverse coordinates that appear in the bulk-to-boundary propagators being integrated. 

\subsection{Probe limit}\label{sec:probe_limit}
 
Here we derive the opening of the wormhole for rotating BTZ using the formula \eqref{eqC} in the probe limit. The first step is to derive the scattering amplitude $e^{i\tilde{\delta}}$, which can be found by studying particles propagating along the horizon whose effect is to produce a shock wave geometry \cite{Shenker:2014cwa}. For simplicity, let us assume the matter creating the shock wave is symmetrically distributed over the transverse space, so that the scattering amplitude will be independent of the transverse coordinates.

The parameters \eqref{eq:a0} for the rotating BTZ metric \eqref{eq:metric_kruskal} are identified as
\begin{equation}
 a_0=\gamma^2,\qquad r_0=r_+,
\end{equation}
with $\gamma$ defined as in \eqref{gamma}. If we send the matter with momentum $p_v$ along $v=0$, this corresponds to a stress tensor\footnote{Upper and lower indices are related by $q^v=-\frac{2}{a_0}q_u$, $p^u=-\frac{2}{a_0}p_v$.}
\begin{equation}
 T_{vv}=\frac{-p_v}{r_+ L}\delta(v), \quad \text{so that} \quad -p_v=\int r_+dx\,dvT_{vv},
\end{equation}
where $L$ denotes the integral of the transverse coordinate without the measure factor $r_+$. For rotating BTZ we have $L=2\pi$, but we keep this factor general to make the dependence on the volume of the transverse space explicit. From the linearized Einstein equations \eqref{linearized} for the $vv$ component, we have
\begin{equation}
 8\pi G_N T_{vv}=\frac{\kappa}{2r_+}h_{vv},
\end{equation}
which gives
\begin{equation}
h_{vv}=a_0 a^u \delta(v), \quad a^u\equiv-\frac{16\pi G_N}{a_0 \kappa L}p_v.
\end{equation}
The effect is then to produce a translation by an amount $a^u$ once we cross the horizon $v=0$. Note that $a^u>0$ since $p_v$ is negative for a physical particle. The scattering amplitude is identified as
\begin{equation} \label{eq:delta}
 i\tilde{\delta}=-ia^uq_u.
\end{equation}

In the probe limit, in which we assume $a^u$ is small, we can expand the exponent in \eqref{eqD} for small $a^u$. The zeroth order term of the expansion cancels with $e^{i\vev{\mathcal{V}}}$ in the expression for $C$ \eqref{eqC}, while the term proportional to $a^u$ contributes to a term that acts as a translation for the $\phi$ wavefunction. In the end we obtain a correlator for $\phi$ that has the form
\begin{equation}
 C_\text{probe}=\vev{\phi_R e^{-ia^v\hat{P}_v}\phi_L},
\end{equation}
where
\begin{equation} \label{eq:av}
 a^v=-\frac{i\alpha}{p^u}\int dq^v\,dx \int dt_1\,dx_1\,q^v \,\tilde{\delta}\,h(t_1,x_1)\Psi_{\mathcal{O}_R}^*(q^v,x)\Psi_{\mathcal{O}_L}(q^v,x),
\end{equation}
and $\hat{P}_v$ is the translation operator that shifts the $\phi_L$ wavefunction by an amount $a^v$ along the $v$ direction. This gives us the opening of the wormhole in the probe limit that was calculated using a different method in Sec.\,\ref{sec:wormhole_size}. The quantity $a^v$ here corresponds to $\Delta V$ in \eqref{eq:DeltaV}.

Let us evaluate \eqref{eq:av} for the rotating BTZ. The wavefunctions $\Psi_{\mathcal{O}_{L/R}}$ can be computed explicitly since we know the form of the bulk-to-boundary propagator, which evaluated at $u=0$ is
\begin{equation}
 \mathcal{K}(0,v,x;t_1,x_1)=\frac{\left(r_+^2-r_-^2\right)^\frac{\Delta}{2}}{2^{\Delta+1}\pi}\left(\frac{1}{-v \frac{\gamma}{2} e^{-r_-(x-x_1)}e^{-\kappa t_1}+\cosh[r_+(x-x_1)]}\right)^\Delta.
\end{equation}
By performing the Fourier transform we obtain the wavefunctions
\begin{align}
 & \Psi_{\mathcal{O}_L}(q^v,x)=\tfrac{2^{1+\Delta} q^v (q^v\gamma)^{-2+\Delta}(r_+\kappa)^{\Delta/2}}{\Gamma(\Delta)}e^{i\cosh[r_+(x-x_1)]e^{-r_-(x-x_1)\kappa t_1}q^v\gamma-\frac{i\pi\Delta}{2}-r_-\Delta(x-x_1)+\kappa t_1},\nonumber\\
 & \Psi^*_{\mathcal{O}_R}(q^v,x)=\Psi_{\mathcal{O}_L}(q^v,x)|_{t_1\to-t_1}.
\end{align}
We can evaluate \eqref{eq:av} by performing the integral over $q^v$ first. The integral over $x_1$ is extended over all real axis because we are summing over the contribution of all images coming from the periodicity in the transverse coordinate. At the end we obtain
\begin{equation} 
 a^v=-\frac{h\, G_N\, 4^{1-2\Delta}r_+^{1+\Delta}\kappa^{1-\Delta}\Gamma(1+2\Delta)}{\gamma\, \Gamma(\Delta)}I_{t_1}I_{x_1},
\end{equation}
where the integrals $I_{t_1}$ and $I_{x_1}$ can be evaluated in terms of beta and hypergeometric functions
\begin{align}
  I_{t_1} &=\int_{t_0}^{t_f} \frac{dt_1}{\cosh^{1+2\Delta}(\kappa\,t_1)}=\frac{i}{2\kappa}\left[B(\cosh^2(\kappa\,t_f),-\Delta,\tfrac{1}{2})-B(\cosh^2(\kappa\,t_0),-\Delta,\tfrac{1}{2})\right], \\ \nonumber 
  I_{x_1} &=\int_{-\infty}^{\infty}\frac{dx\, e^{r_-(x-x_1)}}{\cosh^{1+2\Delta}[r_+(x-x_1)]}\nonumber\\
  & = 2^{1+2\Delta}\left[\tfrac{{}_2F_1\left(\frac{r_++r_-}{2r_+}+\Delta,1+2\Delta,\frac{3}{2}+\frac{r_-}{2r_+}+\Delta,-1\right)}{r_+(1+2\Delta)+r_-}+\tfrac{{}_2F_1\left(\frac{r_+-r_-}{2r_+}+\Delta,1+2\Delta,\frac{3}{2}-\frac{r_-}{2r_+}+\Delta,-1\right)}{r_+(1+2\Delta)-r_-}\right].
\end{align}
This tells us that the $\phi_L$ signal, in the probe limit, is shifted by an amount $a^v$ in the $v$ direction. If we choose $h$ to be positive, then $a^v$ is negative and the wormhole is traversable. The result for the opening agrees with the numerical calculation using the method of \cite{Gao:2016bin} discussed in Sec.\,\ref{sec:wormhole_size}. 

\subsection{Bound from probe limit}\label{sec:bound_from_probe}

In the probe limit we have assumed that the typical momentum of the state created by $\phi$ is small. Once we increase the momentum, the wormhole starts to close and therefore a bound on the information transferred is expected. In this section, following  \cite{Maldacena:2017axo}, and using the opening in probe limit obtained in Sec.\,\ref{sec:probe_limit}, we find a bound on the information that can be transferred through the wormhole. In the next section we present a refined version of the bound using a more careful treatment of the backreaction effects.

Let us assume that we are sending many particles with same characteristic momentum $p_v$. Effectively, this corresponds to treat the wavefunctions $\Psi_{\phi_{L/R}}$ in \eqref{eqC} as multiparticle wavefunctions, and replace the momentum that appear in the scattering amplitude \eqref{eq:delta} by the total momentum $p_v^\text{total}=\sum_l p_v^l$, where $l$ is an index for each individual particle. The amount of information $N_{\text{send}}$ sent through the wormhole can be defined via
\begin{equation}
 N_{\text{send}}\equiv \frac{p_v^{\text{total}}}{p_v}.
\end{equation}

Using the uncertainty principle $p_v\,a^v\gtrsim 1$, we obtain\footnote{A more rigorous derivation using monotonicity of relative entropy instead of the uncertainty principle was also presented in \cite{Maldacena:2017axo} using an argument from \cite{Faulkner:2016mzt}, resulting in the same inequality up to a factor of $2\pi$.}
\begin{equation} \label{eq:Nsend}
 N_\text{send}\lesssim a^vp_v^\text{total}.
\end{equation}
The validity of the probe approximation breaks down when the scattering amplitude $\tilde{\delta}$ becomes order one, so we will require that $\tilde\delta\lesssim 1$. Using \eqref{eq:delta} we obtain
\begin{equation} \label{bound_s}
N_\text{send}\lesssim \frac{|a^v| L \gamma^2\kappa}{16\pi G_N q_u},
\end{equation}
where $q_u$ in the above expression corresponds to the characteristic value of momenta for the particles created by $\mathcal{O}$. 

In the case of $AdS_2$ studied in \cite{Maldacena:2017axo}, the resulting bound (up to an order one constant) was simply given by the value of the coupling. The same happens in our case since the opening of the wormhole $a^v$ is proportional to $h$. The physical interpretation of the bound \eqref{bound_s} becomes more evident by making use of the linearized Einstein equations \eqref{linearized} allowing us to express the bound in terms of the ANE
\begin{equation} \label{eq:boundTUU}
 N_\text{send}\lesssim \frac{r_+L \left|\int du T_{uu}\right|}{q_u}.
\end{equation}
If we assume that the characteristic momenta $q_u$ is order one, we conclude that the bound on information is simply given by the ANE multiplied by the volume of the transverse space. 

\begin{figure}
\center
\includegraphics[width=0.8\linewidth]{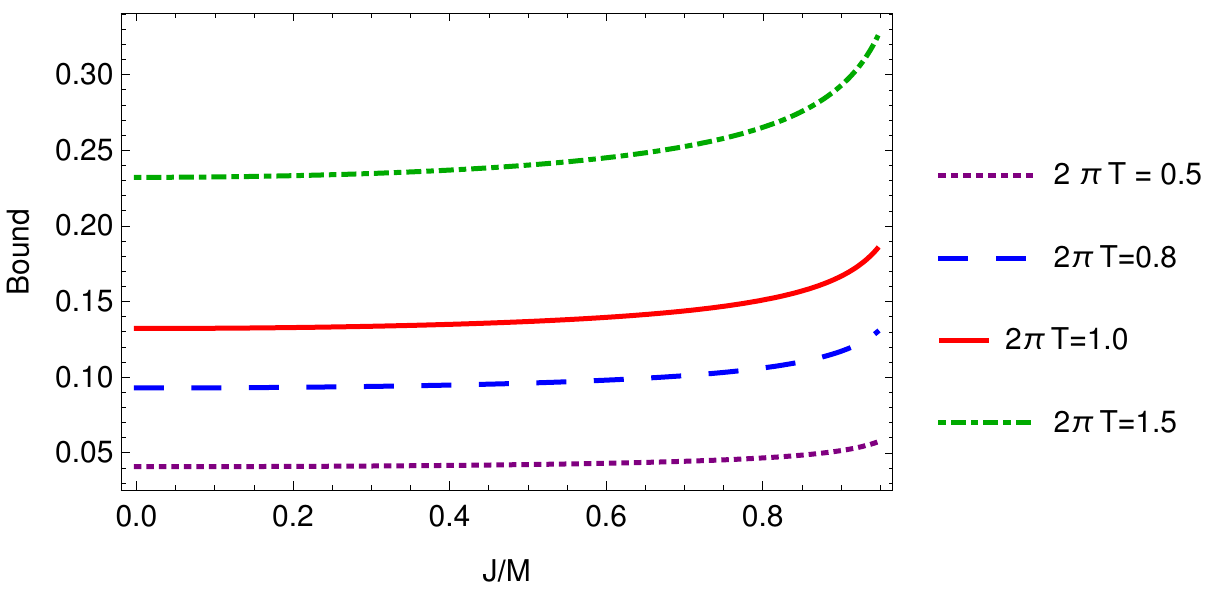}
\caption{Bound on $N_\text{send}$ using Eq. \eqref{bound_s}, assuming momentum is sent spherically over transverse space at fixed temperatures. We fixed $h=1$, $\Delta=0.6$, $t_0=0$, $t_f=1$.}
\label{boundJM}
\end{figure}

There is still an order one constant ambiguity in the above bound. However, the bound still captures the dependence on the temperature and the angular momentum since we carried the dependence on the $r_+$ and $r_-$ parameters in all steps. Fig.\,\ref{boundJM} shows that the bound becomes larger as we increase the dimensionless ratio $J/M$ at fixed temperatures. By looking at \eqref{eq:boundTUU} and from our result for the ANE obtained in Sec.\,\ref{sec:numerics}, we see that this increase is mostly due to a respective increase in the $r_+$ parameter, which is equivalent to an increase in the black hole entropy \eqref{eq:thermo}. 

\begin{figure}[H]
\center
\subfloat[Fixed $M=1$.]{\includegraphics[width=0.45\linewidth]{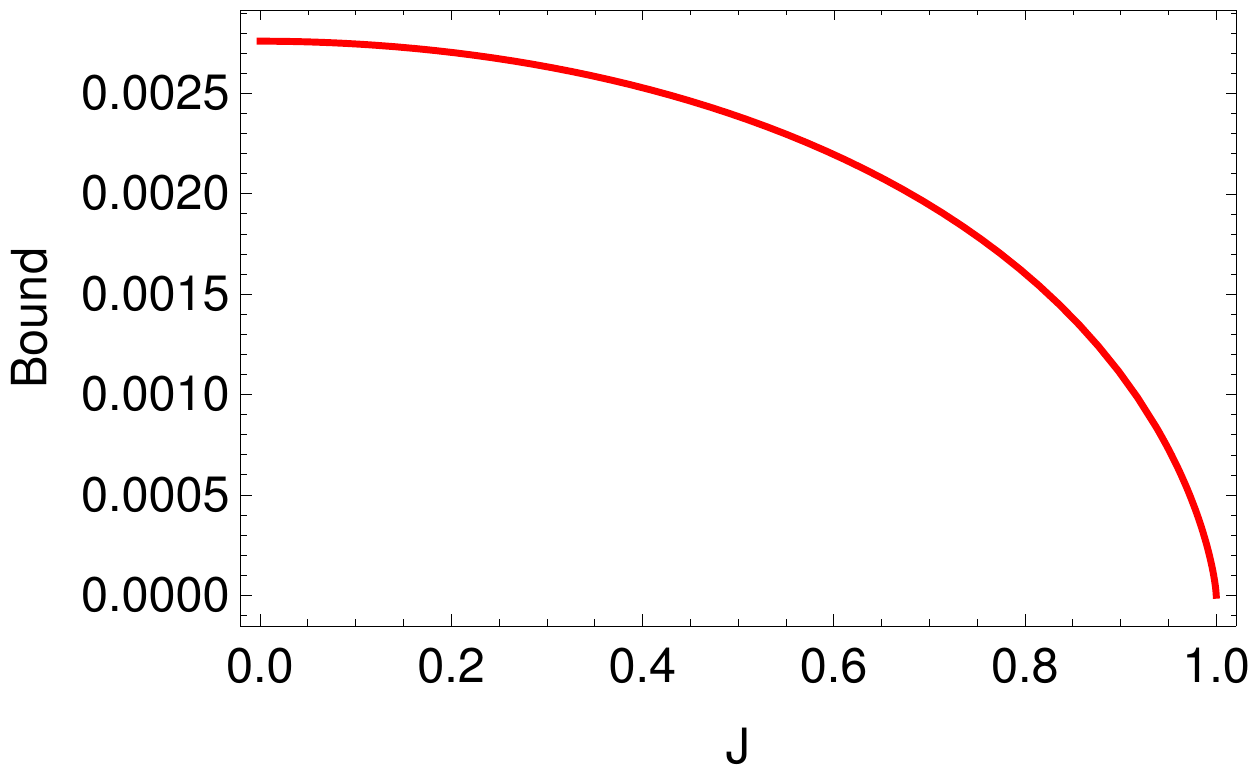}}
\quad
\subfloat[Fixed $r_+=1$.]{\includegraphics[width=0.45\linewidth]{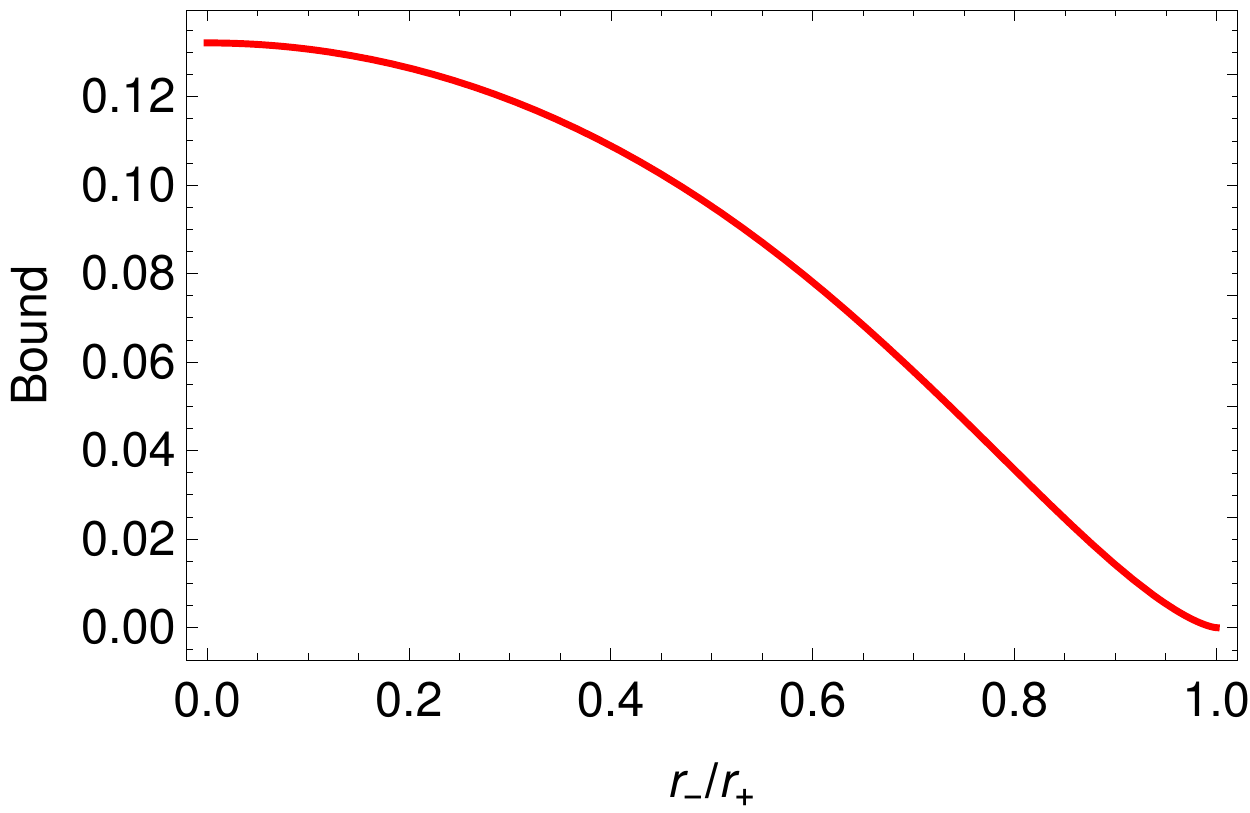}}
\caption{Bound on information transfer.}
\label{bound_extremal} 
\end{figure}

In Fig.\,\ref{bound_extremal}, we show the behavior of the bound as we vary both $J$ and $T$. In particular, as we approach the extremal limit, the bound on information goes to zero, which is consistent with our result in Sec.\,\ref{sec:numerics} where the wormhole closes in the extremal limit. The closing of the wormhole is somewhat  expected since the wormhole throat becomes infinitely long at extremality \cite{Andrade:2013rra, Leichenauer:2014nxa}. We note, however, that the traversable wormhole presented here depends on  several parameters  in the interaction profile \eqref{interaction} such as the time interval in which the interaction is turned on. It is plausible that  scaling the parameters one can  obtain a non-zero opening size even in the extremal limit. Indeed, an example of a traversable wormhole at extremality was recently found in \cite{Fu:2018oaq} by incorporating the backreaction of quantum fields with adequate boundary conditions.

\subsection{Improving analysis with backreaction}\label{sec:improved_bound}

\begin{figure}[h!]
	\center
	\includegraphics[width=0.6\linewidth]{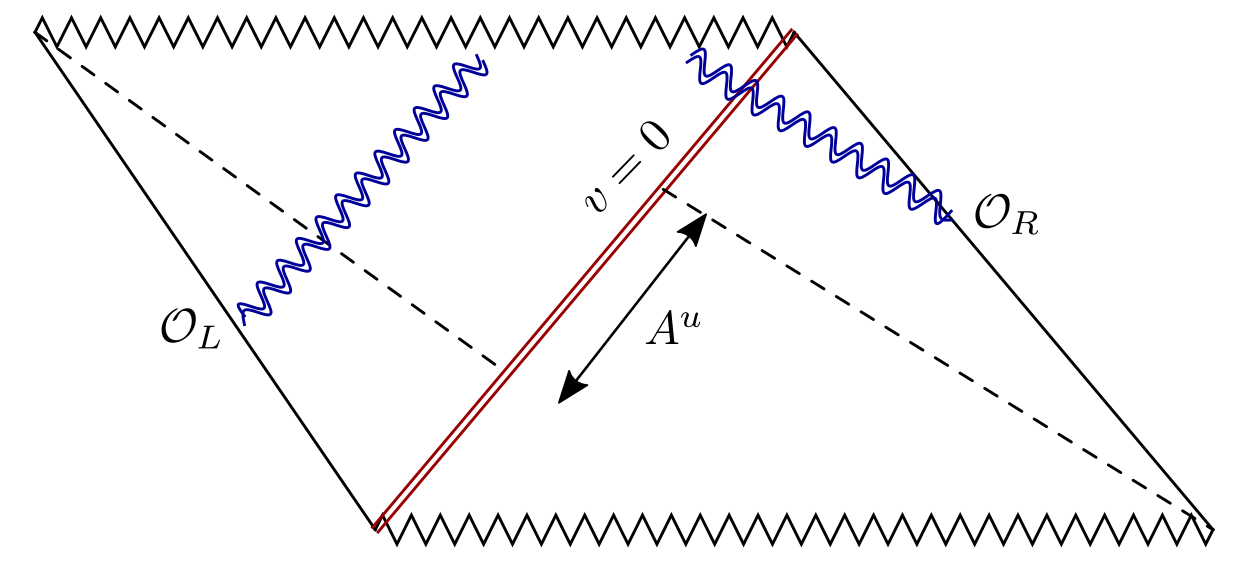}
	\caption{The backreaction of $\phi$ produces a shock wave geometry. The wavefunction associated to $\mathcal{O}_R$ suffers a time delay when it crosses the horizon at $v=0$.}
	\label{fig:setup2}
\end{figure}

Even though the insertion of the operator $\phi$ contributes only to the $T_{vv}$ component of the stress tensor, the ANE along $v=0$ evaluated from \eqref{eq:TUU_pert} still gets modified once we include the backreaction of $\phi$. This can be understood as $\phi$ producing a shock wave localized along $v=0$, which delays the quanta produced by $\mathcal{O}_R$; Fig.\,\ref{fig:setup2}. As a result, the effect of the double trace interaction becomes weaker, decreasing the ANE injected into the bulk and closing the wormhole.

Here, we investigate the effect of backreaction in the opening of the wormhole by applying the double trace deformation in the BTZ geometry perturbed by the shock wave produced by $\phi$. This can be accomplished simply by replacing $U'\to U'+A^u$ in \eqref{eq:pert_prop}, which corresponds to apply a shift in the $U$ direction to the bulk-to-boundary propagator that originally connected the left boundary to a point at the horizon. The parameter $A^u$ is related to the total momentum sent through the wormhole via
\begin{equation}
  A^u=-\frac{16\pi G_N}{a_0 \kappa L}p_v^{\text{total}}.
\end{equation}
Alternatively, using the diagnose of traversability of \cite{Maldacena:2017axo}, we can apply a translation in the wavefunction $\Psi_{\mathcal{O}_R}$
\begin{equation}
 \Psi_{\mathcal{O}_R}\to e^{-iA^u q_u} \Psi_{\mathcal{O}_R}.
\end{equation}

For the purpose of the calculation, we can imagine that we send an extra particle with momentum $p_v\ll p_v^\text{total}$. We can then treat this particle as a probe for the opening of the wormhole that will now depend on $p_v^\text{total}$. In this picture the effective scattering amplitude becomes
\begin{equation}
 i\delta_\text{total}=-i(A^u+a^u)q_u
\end{equation}
where 
\begin{equation}
 a^u=-\frac{16\pi G_N}{a_0 \kappa L}p_v.
\end{equation}
We can expand in $a^u$ to linear order, so that the particle with momentum $p_v$ will acquire a shift in the $v$ direction, which we identify as the opening of the wormhole including the backreaction. The integral over $q^v$ for the exponent of \eqref{eqC} gives
\begin{equation}
 \mathcal{D}=\mathcal{D}_0\int dt_1 dx_1 dx\left(\frac{e^{r_-(x-x_1)}}{\gamma (A^u+a^u)+4 e^{-r_-(x-x_1)}\cosh[r_+(x-x_1)]\cosh(\kappa t_1)}\right)^{2\Delta},
\end{equation}
where the coefficient $\mathcal{D}_0$ is 
\begin{equation}
\mathcal{D}_0=\frac{h r_+^{1+\Delta}\kappa^{2-\Delta}\Gamma(2\Delta)}{\pi\Gamma(\Delta)^2}.
\end{equation}
After expanding to first order in $a^u$ we obtain wormhole opening,
\begin{equation}\label{eq:opening_back}
 a^v_{\text{back}}= \mathcal{A}_0\int \frac{dx\,dt_1\,e^{r_-(x-x_1)}}{(4\pi G_N e^{r_-(x-x_1)}(-p_v^\text{total}/L) +\gamma\kappa\cosh(r_+(x-x_1))\cosh(\kappa t_1))^{1+2\Delta}},
\end{equation}
where $\mathcal{A}_0$ is, 
\begin{equation}\mathcal{A}_0=-4^{1-2\Delta}h G_N \gamma^{2\Delta}r_+^{1+\Delta}\kappa^{2+\Delta}\frac{\Gamma(1+2\Delta)}{\Gamma(\Delta)^2}.
\end{equation}

\begin{figure}[h!]
\center
\includegraphics[width=0.8\linewidth]{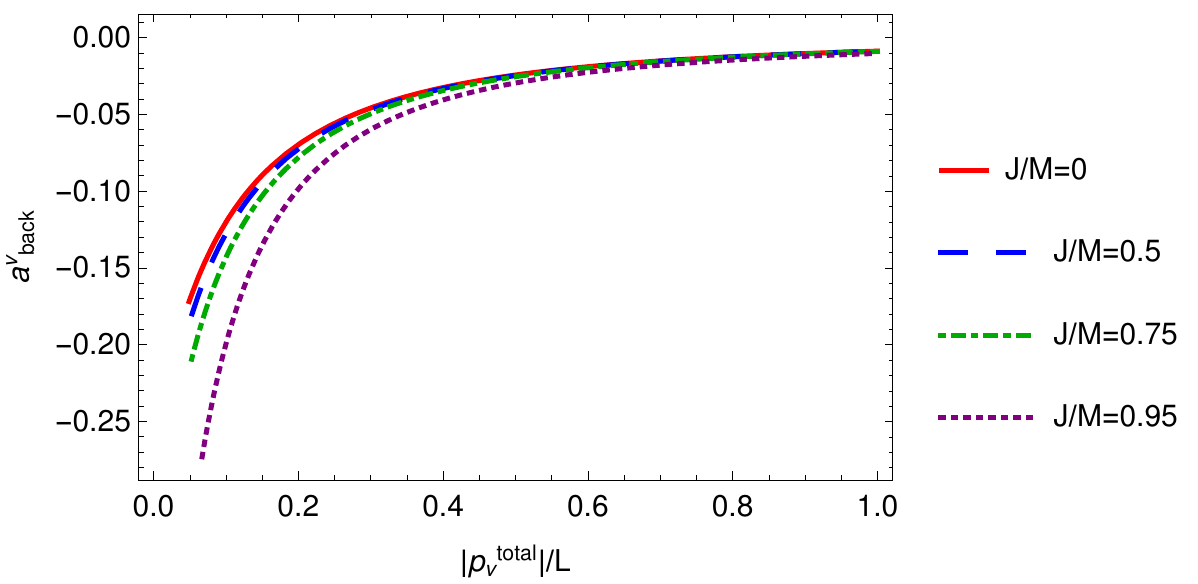}
\qquad
\caption{Opening of the wormhole as a function of momentum $p_v^{\text{total}}$ sent spherically distributed over transverse space. Fixed temperature $T=\frac{1}{2\pi}$.}
\label{fig:openingb_spherical}
\end{figure}

Evaluating \eqref{eq:opening_back} numerically we obtain the opening of the wormhole with backreaction; see Fig.\,\ref{fig:openingb_spherical}. Our parameters were fixed as $h=1$, $G_N=1$, $\Delta=0.6$, $t_0=0$, $t_f=1$, and $T=\frac{1}{2\pi}$. We see that the wormhole closes as we increase the momentum $p_v^{\text{total}}$ sent through it.

\begin{figure}[h!]
\center
\includegraphics[width=0.8\linewidth]{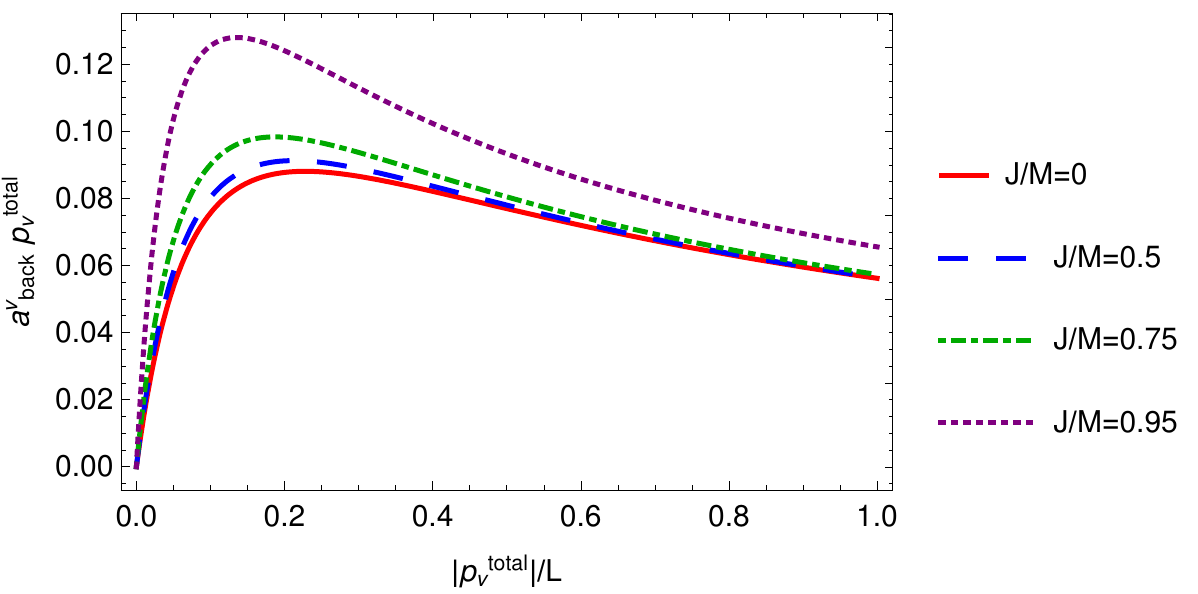}
\qquad
\caption{Bound on information given by $N_\text{send}\lesssim a^v_\text{back}\,p_v^\text{total}$ for spherically distributed momentum. Fixed temperature $T=\frac{1}{2\pi}$.}
\label{fig:boundb_spherical}
\end{figure}

We can obtain a refined bound by determining the momentum that maximizes the right hand side of \eqref{eq:Nsend},
\begin{equation}
 N_\text{send}\lesssim \text{max}\left[a^v_\text{back}\,p_v^\text{total}\right],
\end{equation}
which corresponds to determine the maximum values in the curves displayed in Fig\,\ref{fig:boundb_spherical}. 
This improves the bound by removing the order one ambiguity in the step in which we assumed $\tilde{\delta}\lesssim 1$. Comparing with the previous estimative in Fig.\,\ref{boundJM} at same temperature $T=\frac{1}{2\pi}$, we see that the refined bound is slightly smaller, suggesting a sharper bound.

\section{Dependence on transverse coordinates}\label{sec:transverse}

\begin{figure}[h!]
\center
\includegraphics[width=0.8\linewidth]{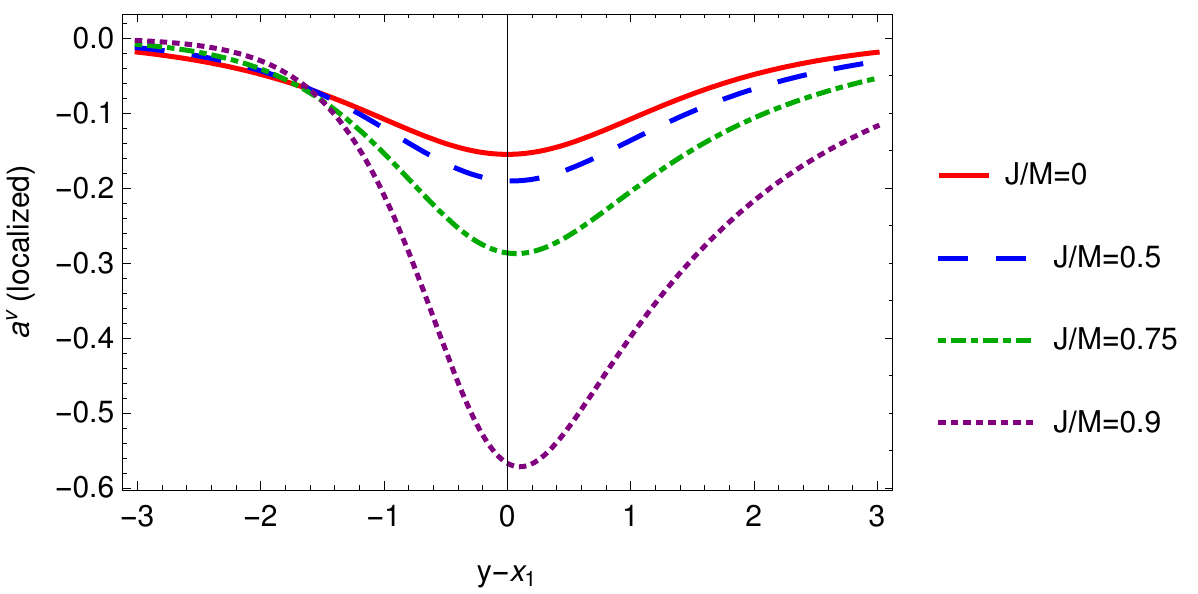}
\qquad
\caption{Opening of the wormhole when interaction is turned only along the $x_1$ direction. Temperature fixed $T=\frac{1}{2\pi}$. Set $g=1$, $\Delta=0.6$, $t_0=0$, and $t_f=1$.}
\label{fig:opening_transverse}
\end{figure}

In the previous sections we have considered the double trace interaction \eqref{interaction} with a uniform coupling so that the size of the opening of the wormhole was independent of the transverse coordinate $x$. A natural question to ask is how the opening changes if we consider instead a non-homogeneous coupling. In this section, we briefly analyze this question by deriving the opening of the wormhole in the probe limit by considering the modified interaction
\begin{equation}
\delta H_\text{loc}(t_1)=- g\,\mathcal{O}_R(t_1,x_1)\mathcal{O}_L(-t_1,x_1),
\end{equation}
which is turned on only in the direction $x_1$ in transverse space, for a time $t_0\leq t_1\leq t_f$. We will follow again the approach used in \cite{Maldacena:2017axo}. We can probe the opening of the wormhole by assuming that the momentum $p_v$ produced by the operator $\phi$ is localized at some position $y_1$. For the background in the form of \eqref{eq:metric_generic}, this contributes to the stress tensor with \cite{Shenker:2014cwa} 
\begin{equation}
 T_{vv}=\frac{a_0}{2 r_0}p^u\delta(y-y_1),
\end{equation}
and the effect is to produce a shock wave localized around $y_1$
\begin{equation}
 h_{vv}(y-y_1)=8\pi G_N r_+ p^u a_0\delta(V)f(y-y_1),
\end{equation}
where $f(y)$ is a transverse profile that can be determined from the linearized Einstein equations. 
Specializing to our rotating geometry the solution is given by
\begin{equation}
 f(y)=\frac{e^{r_-\,y-r_+|y|}}{2r_+},
\end{equation}
which is obtained by solving the equation
\begin{equation}
 -f''(y)+2r_-f'(y)+(r_+^2-r_-^2)f(y)=\delta(y).
\end{equation}
The corresponding scattering amplitude is
\begin{equation} \label{eq:delta_localized}
 \tilde{\delta}_{\text{loc}}=\frac{16\pi G_N r_+ p_v f(x)}{a_0}q_u
\end{equation}
Using this scattering amplitude, we can derive the opening of the wormhole using \eqref{eq:av} again, but now the opening will be a function of the separation $y-x_1$ between $\phi$ and $\mathcal{O}$ in the transverse space. We plotted the result in Fig.\,\ref{fig:opening_transverse}. The wormhole opening is peaked near $x_1$, but due to the presence of rotation the maximum value of the opening is slightly shifted to the right.

\section{Conclusions and future directions}\label{sec:conclusions}
In this work we explored the effect of rotation in the size of a traversable wormhole obtained via a double trace boundary deformation. We find that  the size of the wormhole and the amount of information that can be transferred through it  increases in a rotating geometry.  We improved on the existing bound on information transferred  by taking into account the backreaction.  We  also briefly consider a boundary coupling that has a compact support on the spatial boundary coordinates. We show that this boundary  profile of the coupling  is reflected in the wormhole opening. There are many  issues that remain to be explored in this fascinating subject:  
\begin{itemize}
	\item {\em Charge and extremality.} Given that our results show the rotation tends to make the wormhole larger, it is natural to ask what is the effect of charge in the size of the wormhole. A more detailed study either in the rotating or charged case would also be desirable. 
	\item{\em Traversability in multi-boundary black holes.} In three dimensions multi-boundary black holes are constructed as an appropriate quotient  of AdS. Their causal structure, entanglement entropy and complexity have been thoroughly studied \cite{Aminneborg:1997pz,Balasubramanian:2014hda,Balasubramanian:2016sro,Fu:2018kcp}. It would be interesting to understand how traversability via a boundary deformation works in this scenario and to study its quantum teleportation interpretation. 
	\item {\em Beyond the eikonal approximation.} Both \cite{Gao:2016bin} and \cite{Maldacena:2017axo} rely on the eikonal approximation \cite{Verlinde:1991iu,Kabat:1992tb}. It would be interesting to investigate traversability by performing a  scattering calculation that does not assume this approximation. 
	\item {\em Testing reconstruction behind the horizon.} The traversable wormhole studied here implies a particular deformation of the  boundary Hamiltonian. The effect of this deformation is to bring in causal contact the interior of the black hole and the boundary. Thus, standard  reconstruction methods can now be used for operators that were in the black hole interior and this would allow to check the proposal of \cite{Almheiri:2017fbd} for  operators behind the horizon.  
	\item {\em Higher dimensional wormholes.} Traversable wormholes obtained via double trace deformation have been studied so far only in $1+1$ and $2+1$ dimensions. It would be nice to investigate what changes when we consider traversable wormholes in higher dimensions\footnote{See the recent paper \cite{Maldacena:2018gjk}} in particular how the additional transverse directions can play a role from the boundary perspective as a quantum teleportation. 
\end{itemize}
	We hope to  return to some of  these issues in the future.

\acknowledgments{It is a pleasure to thank Jacques Distler and Don Marolf  for enlightening conversations. E.C. and A.M. thank the ICTP-SAIFR for hospitality and  the participants of the Latin American Workshop in Holography and Gravity (LAWGH) for stimulating discussions. This material is based upon work supported by the National Science Foundation under
	Grant Number PHY-1620610.

	\bibliography{wormhole_ref}{}

\providecommand{\href}[2]{#2}\begingroup\raggedright\begin{thebibliography}{10}

\bibitem{Graham:2007va}
N.~Graham and K.~D. Olum, \emph{{Achronal averaged null energy condition}},
  \href{https://doi.org/10.1103/PhysRevD.76.064001}{\emph{Phys. Rev.}
  {\bfseries D76} (2007) 064001}
  [\href{https://arxiv.org/abs/0705.3193}{{\ttfamily 0705.3193}}].

\bibitem{Kelly:2014mra}
W.~R. Kelly and A.~C. Wall, \emph{{Holographic proof of the averaged null
  energy condition}}, \href{https://doi.org/10.1103/PhysRevD.90.106003,
  10.1103/PhysRevD.91.069902}{\emph{Phys. Rev.} {\bfseries D90} (2014) 106003}
  [\href{https://arxiv.org/abs/1408.3566}{{\ttfamily 1408.3566}}].

\bibitem{Wall:2009wi}
A.~C. Wall, \emph{{Proving the Achronal Averaged Null Energy Condition from the
  Generalized Second Law}},
  \href{https://doi.org/10.1103/PhysRevD.81.024038}{\emph{Phys. Rev.}
  {\bfseries D81} (2010) 024038}
  [\href{https://arxiv.org/abs/0910.5751}{{\ttfamily 0910.5751}}].

\bibitem{Gao:2016bin}
P.~Gao, D.~L. Jafferis and A.~Wall, \emph{{Traversable Wormholes via a Double
  Trace Deformation}},
  \href{https://doi.org/10.1007/JHEP12(2017)151}{\emph{JHEP} {\bfseries 12}
  (2017) 151} [\href{https://arxiv.org/abs/1608.05687}{{\ttfamily
  1608.05687}}].

\bibitem{Morris:1988tu}
M.~S. Morris, K.~S. Thorne and U.~Yurtsever, \emph{{Wormholes, Time Machines,
  and the Weak Energy Condition}},
  \href{https://doi.org/10.1103/PhysRevLett.61.1446}{\emph{Phys. Rev. Lett.}
  {\bfseries 61} (1988) 1446}.

\bibitem{Maldacena:2017axo}
J.~Maldacena, D.~Stanford and Z.~Yang, \emph{{Diving into traversable
  wormholes}}, \href{https://doi.org/10.1002/prop.201700034}{\emph{Fortsch.
  Phys.} {\bfseries 65} (2017) 1700034}
  [\href{https://arxiv.org/abs/1704.05333}{{\ttfamily 1704.05333}}].

\bibitem{Sachdev:1992fk}
S.~Sachdev and J.~Ye, \emph{{Gapless spin fluid ground state in a random,
  quantum Heisenberg magnet}},
  \href{https://doi.org/10.1103/PhysRevLett.70.3339}{\emph{Phys. Rev. Lett.}
  {\bfseries 70} (1993) 3339}
  [\href{https://arxiv.org/abs/cond-mat/9212030}{{\ttfamily
  cond-mat/9212030}}].

\bibitem{Kitaev:2015}
A.~Kitaev, \emph{{A simple model of quantum holography}},  {KITP strings
  seminar and Entanglement 2015 program,
  http://online.kitp.ucsb.edu/online/entangled15/}.

\bibitem{Czech:2018kvg}
B.~Czech, L.~Lamprou and L.~Susskind, \emph{{Entanglement Holonomies}},
  \href{https://arxiv.org/abs/1807.04276}{{\ttfamily 1807.04276}}.

\bibitem{deBoer:2018ibj}
J.~De~Boer, S.~F. Lokhande, E.~Verlinde, R.~Van~Breukelen and K.~Papadodimas,
  \emph{{On the interior geometry of a typical black hole microstate}},
  \href{https://arxiv.org/abs/1804.10580}{{\ttfamily 1804.10580}}.

\bibitem{Yoshida:2018vly}
B.~Yoshida and N.~Y. Yao, \emph{{Disentangling Scrambling and Decoherence via
  Quantum Teleportation}},  \href{https://arxiv.org/abs/1803.10772}{{\ttfamily
  1803.10772}}.

\bibitem{Yoshida:2017non}
B.~Yoshida and A.~Kitaev, \emph{{Efficient decoding for the Hayden-Preskill
  protocol}},  \href{https://arxiv.org/abs/1710.03363}{{\ttfamily 1710.03363}}.

\bibitem{Almheiri:2018ijj}
A.~Almheiri, A.~Mousatov and M.~Shyani, \emph{{Escaping the Interiors of Pure
  Boundary-State Black Holes}},
  \href{https://arxiv.org/abs/1803.04434}{{\ttfamily 1803.04434}}.

\bibitem{Susskind:2017nto}
L.~Susskind and Y.~Zhao, \emph{{Teleportation Through the Wormhole}},
  \href{https://arxiv.org/abs/1707.04354}{{\ttfamily 1707.04354}}.

\bibitem{vanBreukelen:2017dul}
R.~van Breukelen and K.~Papadodimas, \emph{{Quantum teleportation through
  time-shifted AdS wormholes}},
  \href{https://arxiv.org/abs/1708.09370}{{\ttfamily 1708.09370}}.

\bibitem{Bak:2018txn}
D.~Bak, C.~Kim and S.-H. Yi, \emph{{Bulk View of Teleportation and Traversable
  Wormholes}},  \href{https://arxiv.org/abs/1805.12349}{{\ttfamily
  1805.12349}}.

\bibitem{Miyaji:2018atq}
M.~Miyaji, \emph{{Time Evolution after Double Trace Deformation}},
  \href{https://arxiv.org/abs/1806.10807}{{\ttfamily 1806.10807}}.

\bibitem{Banados:1992gq}
M.~Banados, M.~Henneaux, C.~Teitelboim and J.~Zanelli, \emph{{Geometry of the
  (2+1) black hole}}, \href{https://doi.org/10.1103/PhysRevD.48.1506,
  10.1103/PhysRevD.88.069902}{\emph{Phys. Rev.} {\bfseries D48} (1993) 1506}
  [\href{https://arxiv.org/abs/gr-qc/9302012}{{\ttfamily gr-qc/9302012}}].

\bibitem{Banados:1992wn}
M.~Banados, C.~Teitelboim and J.~Zanelli, \emph{{The Black hole in
  three-dimensional space-time}},
  \href{https://doi.org/10.1103/PhysRevLett.69.1849}{\emph{Phys. Rev. Lett.}
  {\bfseries 69} (1992) 1849}
  [\href{https://arxiv.org/abs/hep-th/9204099}{{\ttfamily hep-th/9204099}}].

\bibitem{Hemming:2002kd}
S.~Hemming, E.~Keski-Vakkuri and P.~Kraus, \emph{{Strings in the extended BTZ
  space-time}},
  \href{https://doi.org/10.1088/1126-6708/2002/10/006}{\emph{JHEP} {\bfseries
  10} (2002) 006} [\href{https://arxiv.org/abs/hep-th/0208003}{{\ttfamily
  hep-th/0208003}}].

\bibitem{Balasubramanian:2004zu}
V.~Balasubramanian and T.~S. Levi, \emph{{Beyond the veil: Inner horizon
  instability and holography}},
  \href{https://doi.org/10.1103/PhysRevD.70.106005}{\emph{Phys. Rev.}
  {\bfseries D70} (2004) 106005}
  [\href{https://arxiv.org/abs/hep-th/0405048}{{\ttfamily hep-th/0405048}}].

\bibitem{Maldacena:2001kr}
J.~M. Maldacena, \emph{{Eternal black holes in anti-de Sitter}},
  \href{https://doi.org/10.1088/1126-6708/2003/04/021}{\emph{JHEP} {\bfseries
  04} (2003) 021} [\href{https://arxiv.org/abs/hep-th/0106112}{{\ttfamily
  hep-th/0106112}}].

\bibitem{Krishnan:2009kj}
C.~Krishnan, \emph{{Tomograms of Spinning Black Holes}},
  \href{https://doi.org/10.1103/PhysRevD.80.126014}{\emph{Phys. Rev.}
  {\bfseries D80} (2009) 126014}
  [\href{https://arxiv.org/abs/0911.0597}{{\ttfamily 0911.0597}}].

\bibitem{Levi:2003cx}
T.~S. Levi and S.~F. Ross, \emph{{Holography beyond the horizon and cosmic
  censorship}}, \href{https://doi.org/10.1103/PhysRevD.68.044005}{\emph{Phys.
  Rev.} {\bfseries D68} (2003) 044005}
  [\href{https://arxiv.org/abs/hep-th/0304150}{{\ttfamily hep-th/0304150}}].

\bibitem{Ichinose:1994rg}
I.~Ichinose and Y.~Satoh, \emph{{Entropies of scalar fields on
  three-dimensional black holes}},
  \href{https://doi.org/10.1016/0550-3213(95)00197-Z}{\emph{Nucl. Phys.}
  {\bfseries B447} (1995) 340}
  [\href{https://arxiv.org/abs/hep-th/9412144}{{\ttfamily hep-th/9412144}}].

\bibitem{Azeyanagi:2007bj}
T.~Azeyanagi, T.~Nishioka and T.~Takayanagi, \emph{{Near Extremal Black Hole
  Entropy as Entanglement Entropy via AdS(2)/CFT(1)}},
  \href{https://doi.org/10.1103/PhysRevD.77.064005}{\emph{Phys. Rev.}
  {\bfseries D77} (2008) 064005}
  [\href{https://arxiv.org/abs/0710.2956}{{\ttfamily 0710.2956}}].

\bibitem{Maldacena:2016upp}
J.~Maldacena, D.~Stanford and Z.~Yang, \emph{{Conformal symmetry and its
  breaking in two dimensional Nearly Anti-de-Sitter space}},
  \href{https://doi.org/10.1093/ptep/ptw124}{\emph{PTEP} {\bfseries 2016}
  (2016) 12C104} [\href{https://arxiv.org/abs/1606.01857}{{\ttfamily
  1606.01857}}].

\bibitem{Shenker:2013pqa}
S.~H. Shenker and D.~Stanford, \emph{{Black holes and the butterfly effect}},
  \href{https://doi.org/10.1007/JHEP03(2014)067}{\emph{JHEP} {\bfseries 03}
  (2014) 067} [\href{https://arxiv.org/abs/1306.0622}{{\ttfamily 1306.0622}}].

\bibitem{Shenker:2014cwa}
S.~H. Shenker and D.~Stanford, \emph{{Stringy effects in scrambling}},
  \href{https://doi.org/10.1007/JHEP05(2015)132}{\emph{JHEP} {\bfseries 05}
  (2015) 132} [\href{https://arxiv.org/abs/1412.6087}{{\ttfamily 1412.6087}}].

\bibitem{Verlinde:1991iu}
H.~L. Verlinde and E.~P. Verlinde, \emph{{Scattering at Planckian energies}},
  \href{https://doi.org/10.1016/0550-3213(92)90236-5}{\emph{Nucl. Phys.}
  {\bfseries B371} (1992) 246}
  [\href{https://arxiv.org/abs/hep-th/9110017}{{\ttfamily hep-th/9110017}}].

\bibitem{Kabat:1992tb}
D.~N. Kabat and M.~Ortiz, \emph{{Eikonal quantum gravity and Planckian
  scattering}}, \href{https://doi.org/10.1016/0550-3213(92)90627-N}{\emph{Nucl.
  Phys.} {\bfseries B388} (1992) 570}
  [\href{https://arxiv.org/abs/hep-th/9203082}{{\ttfamily hep-th/9203082}}].

\bibitem{Faulkner:2016mzt}
T.~Faulkner, R.~G. Leigh, O.~Parrikar and H.~Wang, \emph{{Modular Hamiltonians
  for Deformed Half-Spaces and the Averaged Null Energy Condition}},
  \href{https://doi.org/10.1007/JHEP09(2016)038}{\emph{JHEP} {\bfseries 09}
  (2016) 038} [\href{https://arxiv.org/abs/1605.08072}{{\ttfamily
  1605.08072}}].

\bibitem{Andrade:2013rra}
T.~Andrade, S.~Fischetti, D.~Marolf, S.~F. Ross and M.~Rozali,
  \emph{{Entanglement and correlations near extremality: CFTs dual to
  Reissner-Nordström $AdS_5$}},
  \href{https://doi.org/10.1007/JHEP04(2014)023}{\emph{JHEP} {\bfseries 04}
  (2014) 023} [\href{https://arxiv.org/abs/1312.2839}{{\ttfamily 1312.2839}}].

\bibitem{Leichenauer:2014nxa}
S.~Leichenauer, \emph{{Disrupting Entanglement of Black Holes}},
  \href{https://doi.org/10.1103/PhysRevD.90.046009}{\emph{Phys. Rev.}
  {\bfseries D90} (2014) 046009}
  [\href{https://arxiv.org/abs/1405.7365}{{\ttfamily 1405.7365}}].

\bibitem{Fu:2018oaq}
Z.~Fu, B.~Grado-White and D.~Marolf, \emph{{Toward self-supporting wormholes}},
   \href{https://arxiv.org/abs/1807.07917}{{\ttfamily 1807.07917}}.

\bibitem{Aminneborg:1997pz}
S.~Aminneborg, I.~Bengtsson, D.~Brill, S.~Holst and P.~Peldan, \emph{{Black
  holes and wormholes in (2+1)-dimensions}},
  \href{https://doi.org/10.1088/0264-9381/15/3/013}{\emph{Class. Quant. Grav.}
  {\bfseries 15} (1998) 627}
  [\href{https://arxiv.org/abs/gr-qc/9707036}{{\ttfamily gr-qc/9707036}}].

\bibitem{Balasubramanian:2014hda}
V.~Balasubramanian, P.~Hayden, A.~Maloney, D.~Marolf and S.~F. Ross,
  \emph{{Multiboundary Wormholes and Holographic Entanglement}},
  \href{https://doi.org/10.1088/0264-9381/31/18/185015}{\emph{Class. Quant.
  Grav.} {\bfseries 31} (2014) 185015}
  [\href{https://arxiv.org/abs/1406.2663}{{\ttfamily 1406.2663}}].

\bibitem{Balasubramanian:2016sro}
V.~Balasubramanian, J.~R. Fliss, R.~G. Leigh and O.~Parrikar,
  \emph{{Multi-Boundary Entanglement in Chern-Simons Theory and Link
  Invariants}}, \href{https://doi.org/10.1007/JHEP04(2017)061}{\emph{JHEP}
  {\bfseries 04} (2017) 061}
  [\href{https://arxiv.org/abs/1611.05460}{{\ttfamily 1611.05460}}].

\bibitem{Fu:2018kcp}
Z.~Fu, A.~Maloney, D.~Marolf, H.~Maxfield and Z.~Wang, \emph{{Holographic
  complexity is nonlocal}},
  \href{https://doi.org/10.1007/JHEP02(2018)072}{\emph{JHEP} {\bfseries 02}
  (2018) 072} [\href{https://arxiv.org/abs/1801.01137}{{\ttfamily
  1801.01137}}].

\bibitem{Almheiri:2017fbd}
A.~Almheiri, T.~Anous and A.~Lewkowycz, \emph{{Inside out: meet the operators
  inside the horizon. On bulk reconstruction behind causal horizons}},
  \href{https://doi.org/10.1007/JHEP01(2018)028}{\emph{JHEP} {\bfseries 01}
  (2018) 028} [\href{https://arxiv.org/abs/1707.06622}{{\ttfamily
  1707.06622}}].

\bibitem{Maldacena:2018gjk}
J.~Maldacena, A.~Milekhin and F.~Popov, \emph{{Traversable wormholes in four
  dimensions}},  \href{https://arxiv.org/abs/1807.04726}{{\ttfamily
  1807.04726}}.

\end{thebibliography}\endgroup
\bibliographystyle{JHEP}

\end{document}